\documentclass[11pt]{article}

\usepackage{graphicx}
\usepackage{times}
\usepackage{amsmath}
\usepackage{amsfonts}
\usepackage{amssymb}
\usepackage[super,comma,sort&compress]{natbib}
\usepackage[hidelinks,implicit=false,pdfpagelabels=false,pdfpagemode=UseNone]{hyperref}

\newenvironment{supplement}[1]
{%
  \clearpage
  \let\orignewcommand\newcommand
  \let\newcommand\renewcommand
  \makeatletter
  \input{size#1.clo}%
  \makeatother
  \let\newcommand\orignewcommand
	\setlength{\textheight}{23cm} \setlength{\textwidth}{17.5cm}
	\headsep 0mm
	\headheight 0mm
	\setlength{\oddsidemargin}{\dimexpr(\paperwidth-\textwidth)/2-1in}
	\setlength{\evensidemargin}{\oddsidemargin}
	\setlength{\topmargin}{\dimexpr(\paperheight-\textheight)/2-\headheight-\headsep-1in}

	\setcounter{equation}{0}
	\setcounter{figure}{0}
	\setcounter{table}{0}
	\setcounter{page}{1}
	\setlength{\parindent}{0pt}
}

\begin{document}

\makeatletter
\def\@biblabel#1{#1.}
\makeatother

\title{Opto-valleytronic imaging of atomically thin semiconductors}

\author{Andre Neumann$^{1}$, Jessica Lindlau$^{1}$, L\'{e}o Colombier$^{1}$, Manuel Nutz$^{1}$, \\
Sina Najmaei$^{3}$, Jun Lou$^{3}$, Aditya D. Mohite$^{2}$, Hisato
Yamaguchi$^{2}$, and Alexander H\"ogele$^{1}$}

\date{}

\maketitle

\begin{center}
{\small
$^1$Fakult\"at f\"ur Physik, Munich Quantum Center, and Center for NanoScience (CeNS),\\
Ludwig-Maximilians-Universit\"at M\"unchen, D-80539 M\"unchen, Germany\\

$^2$MPA-11 Materials Synthesis and Integrated Devices, Materials Physics and Applications Division,\\
Los Alamos National Laboratory (LANL), Los Alamos, NM 87545, U.S.A.\\

$^3$Department of Materials Science and NanoEngineering, Rice University, Houston, TX 77005, U.S.A.}
\end{center}

\vspace*{\stretch{1}} {\bf Transition metal dichalcogenide semiconductors represent elementary components of layered heterostructures for emergent technologies beyond conventional opto-electronics. In their monolayer form they host electrons with quantized circular motion and associated valley polarization and valley coherence as key elements of opto-valleytronic functionality. Here, we introduce two-dimensional polarimetry as means of direct imaging of the valley pseudospin degree of freedom in monolayer transition metal dichalcogenides. Using MoS$_2$ as a representative material with valley-selective optical transitions, we establish quantitative image analysis for polarimetric maps of extended crystals, and identify valley polarization and valley coherence as sensitive probes of crystalline disorder. Moreover, we find site-dependent thermal and non-thermal regimes of valley-polarized excitons in perpendicular magnetic fields. Finally, we demonstrate the potential of wide-field polarimetry for rapid inspection of opto-valleytronic devices based on atomically thin semiconductors and heterostructures.}
\vspace*{\stretch{1}}

Published in \textit{Nature Nanotechnology} \textbf{12}, 329--334 (2017).

The version of record is available online at \url{https://doi.org/10.1038/nnano.2016.282}.
\clearpage

The valley quantum degree of freedom manifests itself as the quantized angular motion of crystal electrons near the band edges. In analogy to spin it represents a resource for quantum information in conventional semiconductors such as aluminum arsenide \cite{Gunawan2007} and silicon \cite{Culcer2012}, or in atomically thin materials including graphene \cite{Rycerz2007} and TMDs \cite{Xiao2012,Xu2014}. Direct band-gap monolayer TMDs \cite{Splendiani2010,Mak2010} are particularly viable for practical realizations of valleytronic concepts as they enable initialization \cite{Xiao2012}, manipulation \cite{Kim2014} and detection \cite{Cao2012,Mak2012,Zeng2012} of the valley pseudospin by all-optical means. Despite successful realization of first opto-valleytronic devices \cite{Zhang2014,Mak2014}, controversy prevails with respect to possible intrinsic and extrinsic origins of significant variations in the degrees of valley polarization \cite{Cao2012,Mak2012,Zeng2012,Sallen2012,Lagarde2014} and valley coherence \cite{Jones2013,Wang2014} observed for different material representatives of monolayer TMD semiconductors.

The valley pseudospin in TMDs is most conveniently accessed with photoluminescence (PL) polarimetry of band-edge excitons \cite{Xu2014}. The degrees of circular and linear PL polarization, $P_c$ and $P_l$, defined as the ratio of emission intensities $P=(I_{co} - I_{cr})/(I_{co} + I_{cr})$ detected in co-polarized ($I_{co}$) and cross-polarized ($I_{cr}$) configurations with a circularly ($\sigma$) or linearly ($\pi$) polarized excitation laser, are direct measures of valley polarization \cite{Cao2012,Mak2012,Zeng2012} and valley coherence \cite{Jones2013,Wang2014}. Most values reported for circular and linear PL polarizations in monolayer TMDs are well below unity, and they vary significantly with material quality or the underlying substrate \cite{Cao2012,Mak2012,Zeng2012,Sallen2012,Lagarde2014,Wu2013,Jones2014,Zhu2014}. A detailed understanding of the variations in $P_c$ and $P_l$ has been elusive to date and partly attributed to different ratios of the exciton and valley dynamics in different samples. Analogous to optical spin orientation \cite{Meier1984}, ideal initialization of the valley polarization yields in a steady-state PL measurement $P_c=1/(1+\tau_{0}/\tau_{l})$ close to unity if the exciton lifetime $\tau_0$ is short as compared to the longitudinal valley decay time $\tau_l$. The same argument holds for $P_l$ as a measure of valley coherence with the transverse valley decay time $\tau_t$. For TMD monolayers with long-lived excitons, one therefore expects a sizable reduction in the degrees of circular and linear polarizations due to exchange-mediated valley decay and dephasing \cite{Maialle1993,Glazov2014,Yu2014a} active during the exciton lifetime.

In the following we demonstrate that the notion of the degrees of circular and linear polarization as being determined simply by the exciton and valley lifetimes is of limited validity, and therefore previous quantitative conclusions drawn from this simplistic picture should be critically revised. However, the model is helpful for a qualitative interpretation of varying degrees of valley polarization and coherence across single TMD monolayers. A more quantitative analysis can be provided by taking into account the optical valley initialization processes. This insight is based on our experiments on extended MoS$_2$ flakes grown by chemical vapor deposition (CVD) and transferred onto SiO$_2$/Si substrates as a representative TMD material system (see the Supplementary Information for sample details). The valley pseudospin physics in single- and poly-crystalline MoS$_2$ monolayers were addressed with steady-state two-dimensional circular and linear PL polarimetry.

{\bf \large{Photoluminescence spectroscopy of extended monolayers}}

\begin{figure*}[t]
\begin{center}
\includegraphics[scale=1.0]{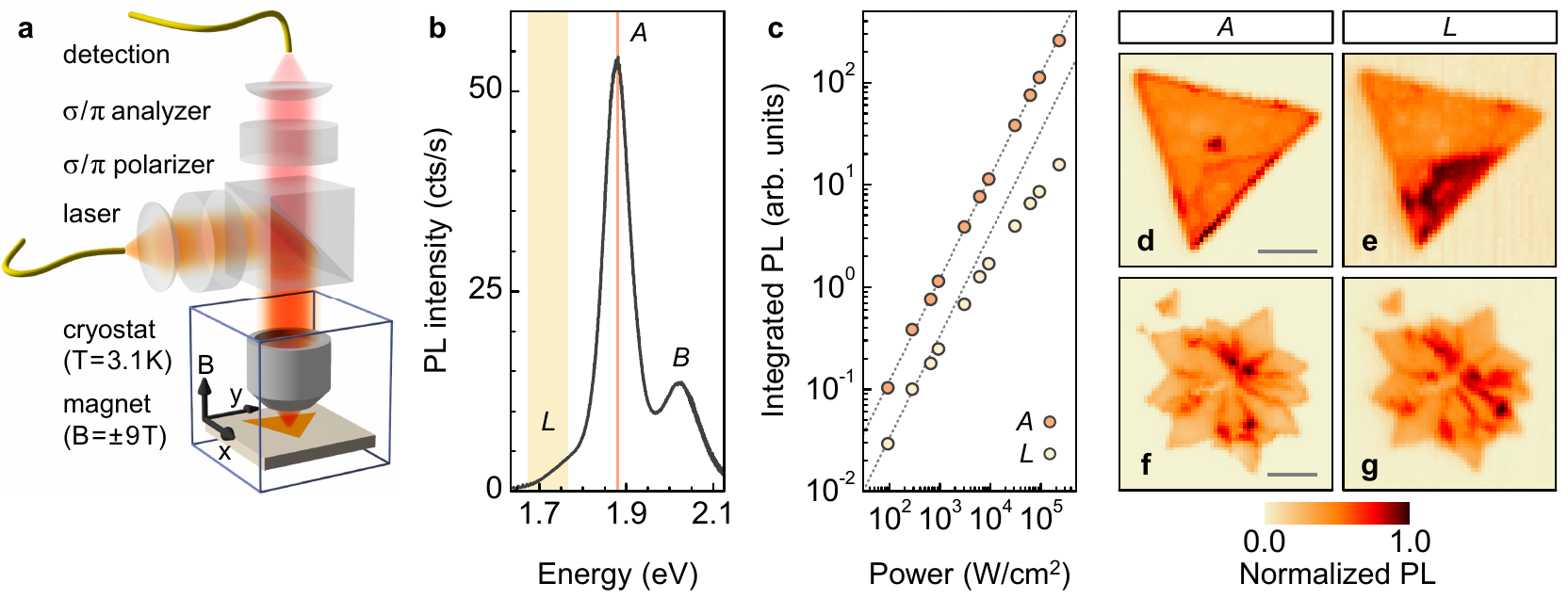}
\caption{\textbf{Confocal spectroscopy and imaging of extended MoS$_2$ monolayers grown by chemical vapor deposition.} \textbf{a}, Experimental setup schematics: the sample with extended monolayer MoS$_2$ crystals on SiO$_2$ is positioned within diffraction-limited confocal laser excitation and photoluminescence detection spots ($0.7~\mu$m diameter) of a low-temperature apochromatic objective in a closed-cycle cryostat with a base temperature of $3.1$~K. A solenoid allows to apply magnetic fields of up to $9$~T perpendicular to the crystal plane. The excitation and detection channels feature polarizing optical components for photoluminescence polarimetry in circular ($\sigma$) and linear ($\pi$) bases. \textbf{b}, Cryogenic photoluminescence spectrum of monolayer MoS$_2$ with $A$, $B$ and low-energy ($L$) exciton features. \textbf{c}, Photoluminescence intensities of $A$ and $L$ excitons as a function of laser power recorded within the colored spectral bands in \textbf{b}; dashed lines indicate linear response. \textbf{d}, \textbf{e}, and \textbf{f}, \textbf{g}, Raster-scan images of the photoluminescence intensity within the $A$ and $L$ exciton bands for a single- and poly-crystalline MoS$_2$ flake, respectively (the scale bars are $5~\mu$m). The data were recorded at $3.1$~K with a laser at $532$~nm.} \label{fig1}
\end{center}
\end{figure*}

We begin our studies by characterizing individual MoS$_2$ crystals with cryogenic PL. The experimental setup for confocal PL spectroscopy and raster-scan imaging is shown schematically in Fig.~\ref{fig1}a. The sample was cooled to $3.1$~K in a closed-cycle cryostat and positioned within the confocal excitation and detection spots of a low-temperature apochromatic objective. The polarization of the excitation and detection pathways was set independently for circular or linear polarimetry by according combinations of linear polarizers, half- and quarter-waveplates. A superconducting solenoid was used to apply magnetic fields up to $9$~T perpendicular to the sample.

A representative cryogenic spectrum of monolayer MoS$_2$ recorded with a non-resonant excitation laser at $532$~nm is shown in Fig.~\ref{fig1}b. It features $A$ and $B$ exciton PL around $1.9$~eV and $2.0$~eV characteristic of monolayer MoS$_2$ on SiO$_2$ \cite{Splendiani2010,Mak2010}. Additionally, we observed a red-shifted PL of localized excitons ($L$) which exhibit saturation as a function of the excitation laser power \cite{Tongay2013} in contrast to $A$ excitons with linear response (Fig.~\ref{fig1}c). The maps in Fig.~\ref{fig1}d~-~g, constructed by color-coding normalized raster-scan PL within exciton-specific bands indicated in Fig.~\ref{fig1}b, show the $A$ and $L$ exciton PL profiles for two representative MoS$_2$ flakes of our studies. The triangular shape (Fig.~\ref{fig1}d and e) is typical for CVD-grown single-crystal flakes, while the star-shaped geometry of the poly-crystalline flake (Fig.~\ref{fig1}f and g) reflects a cluster of single-crystal domains separated by grain boundaries \cite{Zande2013, Najmaei2013}.

The PL intensity profiles show spatial variations across the flakes due to the presence of spatial inhomogeneities in the crystal quality. The edges and the center of the single-crystal triangle exhibit a more intense $A$ exciton PL than the rest of the flake (Fig.~\ref{fig1}d), while the $L$ exciton emission is most intense in a 'puddle' at the right edge of the triangle (Fig.~\ref{fig1}e). Based on previous studies \cite{Zande2013, Najmaei2013}, Raman spectroscopy (see the Supplementary Information) and optical inspection of other flakes on our sample, we identify the central region as a bilayer triangle formed on top of the monolayer single-crystal MoS$_2$. The puddle is likely a collection of point defect contaminants responsible for exciton localization \cite{Tongay2013}. The poly-crystalline star shows homogeneous domains of both $A$ and $L$ emission separated by lines of enhanced intensity (Fig.~\ref{fig1}f and g) at the grain boundaries \cite{Zande2013, Najmaei2013} which also seem to favor exciton localization. At low excitation powers, emission hot-spots of cryogenic quantum dots \cite{Srivastava2015a, He2015, Koperski2015, Chakraborty2015} dominate the PL intensity profiles of $L$ excitons (see the Supplementary Information).

\vspace{13pt} {\bf \large{Raster-scan opto-valleytronic imaging}}

\begin{figure}[t]
\begin{center}
\includegraphics[scale=1.0]{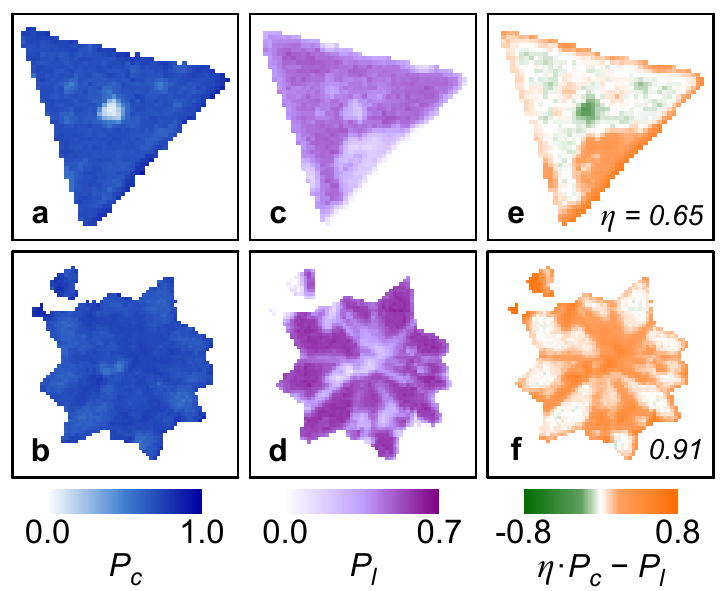}
\caption{\textbf{Raster-scan polarimetry of single- and poly-crystalline MoS$_2$.} \textbf{a}, \textbf{b}, Circular and \textbf{c}, \textbf{d}, linear polarimetric profiles of the photoluminescence within the $A$ exciton band for single- and poly-crystalline MoS$_2$ of Fig.~1, respectively. The false-color maps in \textbf{e} and \textbf{f} were computed as scaled differences of $P_c$ and $P_l$ (with different scaling factors $\eta$ as given in \textbf{e} and \textbf{f}) to highlight the regions of bilayer formation (green) and valley decoherence at crystal defects (orange). All data were recorded with an excitation laser at $637$~nm; the temperature was $3.1$~K.}
\label{fig2}
\end{center}
\end{figure}

The polarization-resolving feature of our setup allows us to identify characteristic signatures of crystal defects with circular and linear polarimetry. The two-dimensional maps of the degrees of circular and linear PL polarizations in Fig.~\ref{fig2}a~-~d were recorded for the flakes of Fig.~\ref{fig1} with a laser at $637$~nm in resonance with the blue shoulder of the $A$ exciton PL which typically results in high near-resonant polarization values \cite{Mak2012,Jones2013,Wang2015,Lagarde2014}. In the absence of an external magnetic field, time-reversal symmetry implies identical degrees of $K$ and $K'$ valley polarization and thus equal $P_c$ recorded with $\sigma^+$ and $\sigma^-$ polarimetry \cite{Xiao2012}. Equivalently, the valley coherence is independent of the choice of the linear basis \cite{Jones2013} (see the Supplementary Information for the equivalence of complementary bases at zero magnetic field). First, we focus our analysis on the fundamental $A$ exciton to discuss the polarization in the $K$ valley, and the valley coherence in terms of $K$-$K'$ superpositions.

The feature of highest contrast in the $P_c$ maps of the single- and poly-crystalline flakes is located at the triangle center (Fig.~\ref{fig2}a) and coincides with the bilayer region discussed earlier. Optical selection rules are different for TMD bilayers \cite{Gong2013} and provide direct means to single out bilayer regions in polarimetric imaging. Away from bilayer regions and edges, the degree of $P_c$ in Fig.~\ref{fig2}a and b is remarkably high and homogeneous. Most surprisingly, the grain boundaries of the poly-crystalline star are almost invisible in the map of Fig.~\ref{fig2}b. This observation of homogeneous valley polarization is in stark contrast to the sizable variations in the PL intensity profiles (Fig.~\ref{fig1}d and f). It is also strongly contrasted by the pronounced inhomogeneities of the respective $P_l$ profiles in Fig.~\ref{fig2}c and d, where both the puddle and the grain boundaries appear most prominently as sites of low valley coherence. To discuss these differences qualitatively, we recall the simple model where valley depolarization and dephasing are responsible for the reduction of $P_c$ and $P_l$ during the exciton lifetime. High $P_c$ values in the maps of Fig.~\ref{fig2}a and b imply that the degree of optical valley pseudospin initialization is high in our experiments -- a necessary condition for the optical generation of valley coherence. With the exception of the bilayer region, the homogenous $P_c$ profiles also suggest that the longitudinal valley relaxation is slow on the timescale of the exciton lifetime even in the presence of disorder. On the same timescale, however, rapid transverse valley relaxation results in locally reduced $P_l$ values as in the defective regions of the flakes in Fig.~\ref{fig2}c and d.

In order to visualize the regions of rapid valley dephasing signified by large differences in $P_c$ and $P_l$, we present differential polarization profiles for both representative flakes. Fig.~\ref{fig2}e and f were computed with a scaling factor $\eta$ of $0.65$ and $0.91$, respectively, to enhance the map contrast and thus the visibility of monolayer defects by highlighting bilayers in green and sites of valley decoherence (crystal edges, grain boundaries and the puddle of surface contaminants) in orange. Orange regions of disorder are characterized by rapid dephasing of optically generated quantum coherent superpositions of $K$ and $K'$ excitons without signatures of low valley polarization. The white areas of the differential profiles in Fig.~\ref{fig2}e and f identify the least defective monolayer crystals most favorable for opto-valleytronic applications.

\clearpage {\bf \large{Opto-valleytronic imaging in magnetic field}}

Our analysis is complemented by circular polarimetric profiling in non-zero magnetic fields. A positive magnetic field oriented orthogonal to the TMD monolayer plane breaks the time-reversal symmetry and lifts the valley degeneracy by decreasing (increasing) the exciton energy in the $K$ ($K'$) valley \cite{Li2014, Srivastava2015, Aivazian2015, MacNeill2015,Wang2015a,Stier2016}. Moreover, it changes the degree of circular polarization \cite{Li2014,Srivastava2015,Aivazian2015,MacNeill2015,Wang2015a}, which we monitored for our flakes with magneto-optical polarimetry. The polarimetric maps recorded in a magnetic field of $+9$~T within the $A$ and $L$ exciton bands are shown in Fig.~\ref{fig3}a~-~d for the representative single-crystal triangle (see the Supplementary Information for complementary measurements in a magnetic field of $-9$~T). The main signature of the valley Zeeman effect in opto-valleytronic imaging is the field-induced difference in $P_c$ measured with $\sigma^+$ and $\sigma^-$ excitations (compare Fig.~\ref{fig3}a and c with b and d). In a positive field of $9$~T the $P_c$ within both $A$ and $L$ spectral bands exhibits a decrease (increase) under $\sigma^+$ ($\sigma^-$) excitation in regular regions of the triangular flake.

To analyze the field-induced changes in the valley polarization of $A$ excitons we removed the contribution of the trion PL from the $A$ exciton band by spectral deconvolution (see the Supplementary Information). The resulting evolution of $P_c$ with magnetic field on a regular site of the single-crystal triangle is presented in the upper and central panels of Fig.~\ref{fig3}e for the deconvolved $A$ exciton and for the $L$ exciton band, respectively. The data in the upper panel of Fig.~\ref{fig3}e exhibit an evolution similar to $A$ excitons in WSe$_2$ monolayer \cite{Aivazian2015} with an $X$-shaped pattern that can be much less pronounced on different samples and under less resonant excitation \cite{Wang2015a}. This $X$-shape is not a generic feature of TMD magnetoluminescence as it is distinct from the $P_c$ evolutions of MoSe$_2$ neutral excitons in magnetic field\cite{MacNeill2015,Wang2015a}. It rather reflects a counter-intuitive steady-state distribution of excitons among the two Zeeman-split $K$ and $K'$ valleys \cite{Aivazian2015}: at positive magnetic fields the population in the energetically higher valley Zeeman branch $K'$ becomes progressively protected from the relaxation into the lower branch $K$ with increasing magnetic field. This 'hot' valley exciton population on a regular position is contrasted by the 'thermal' distribution of $L$ excitons in the puddle (Fig.~\ref{fig3}e, lower panel) with a sign reversal of the $X$-pattern as one would expect for a population redistribution that favors the exciton state of lowest energy.

The scheme of non-thermal, optically induced valley population imbalance is the opto-valleytronic counterpart of optical spin orientation in conventional semiconductors \cite{Meier1984}. For the fundamental $A$ exciton, rate-equation analysis \cite{Aivazian2015} suggests that out-of-equilibrium $K$ and $K'$ valley populations result from both a finite branching of the optically excited valley polarization into valley flipping and conserving relaxation channels and a slow valley depolarization in the exciton ground state. In fact, valley depolarization times much longer than the exciton decay time underpin the non-thermal valley population regime \cite{Aivazian2015}. Our analysis that accounts for inter-valley thermalization in addition to finite branching of the photoexcited population yields best fits to the data (solid lines in Fig.~\ref{fig3}e) with $\tau_l / \tau_0 \simeq 17$ for $A$ and $50$ for $L$ and $L_P$ excitons (see the Supplementary Information for model details and other relevant fit parameters). In contrast, the model with ideal initial polarization of the luminescent exciton states predicts $\tau_l /\tau_0 \simeq 9$ for $A$ and $0.1$ for $L$ excitons, respectively. The discrepancies stem from the ignorance of the idealized model to the details of the valley pseudospin initialization in $K$ and $K'$ exciton ground states. Actually it holds only if the valley conserving relaxation significantly outcompetes the valley flipping relaxation during the formation process of the $K$ and $K'$ excitons out of the selectively excited valley as in the case of $A$ excitons in our experiment with near-resonant excitation. This does not apply to localized states, where significant population branching occurs upon relaxation, and different branching scenarios determine the non-thermal and thermal populations of $L$ and $L_P$ excitons: for the former, alike for $A$ excitons, the valley polarization is increasingly accumulated in the photoexcited upper valley due to a reduction of branching with magnetic field, while for the the latter defect-assisted relaxation likely renders the branching field-independent.

The regime of $\tau_l/ \tau_0 \gg 1$ identified by our analysis for both $A$ and $L$ excitons in monolayer MoS$_2$ yields upper bounds on the longitudinal valley depolarization times of $\tau_l \simeq 80$~ps for $A$ and $230$~ps for $L$ and $L_P$ excitons if we take the PL decay time of $\tau_0 = 4.5$~ps for both the $A$ excitons in MoS$_2$ \cite{Lagarde2014} and the dominant decay timescale of localized excitons in our sample (see the Supplementary Information for time-resolved PL data). In absolute terms, the longitudinal depolarization time of $A$ excitons might be shorter if scaled to a more rapid exciton decay potentially present in our CVD-grown MoS$_2$ flakes but inaccessible in our experiments due to limited temporal resolution. In relative terms, however, our finding of $\tau_l \gg \tau_0$ is in accord with previous results on WSe$_2$ \cite{Aivazian2015}. At the same time it is contrasted by the theoretical estimate $\tau_l \simeq \tau_0 \simeq 1$~ps for the valley depolarization dynamics in the presence of long-range exchange \cite{Glazov2014,Yu2014a} which in turn is an integral part of the finite-branching model \cite{Aivazian2015}. This caveat is qualitatively resolved by the notion of exchange-mediated valley depolarization timescales beyond $1$~ps for luminescent excitons with small center-of-mass momentum away from the light-cone edges \cite{Glazov2014,Yu2014a}. For quantitative consistency, however, efforts in theory and experiment on the details of exciton valley dynamics in the presence of valley conserving and flipping relaxation channels and external magnetic fields are required beyond the scope of this work.

\vspace{13pt} {\bf \large{Wide-field opto-valleytronic imaging}}

We return to polarimetric mapping to demonstrate that the main valleytronic signatures of layered TMDs discussed above can be obtained in direct two-dimensional imaging. To this end we defocussed the excitation laser to illuminate a spot of $\sim 100~\mu$m diameter and replaced the single-mode fiber in the detection path with an imaging lens and a CCD. The resulting optical system images a sample area of $\sim (0.1 \times 0.1)~\mu$m$^2$ onto a single pixel of a standard room-temperature CCD array. A tunable band-pass filter placed before the CCD imaging lens was used to select exciton-specific bands, and polarimetric imaging was performed either in the circular or linear basis.

Polarimetric images of the two representatives flakes recorded in cross-linear configuration (orthogonal excitation and detection polarizations) within the $A$ exciton band are shown in Fig.~\ref{fig4}a and b. In the given configuration, the high intensity features correspond to bilayer and dephasing regions of the flakes. Apart from a scaling factor, the images are equivalent to gray-scale versions of the false-color maps in Fig.~\ref{fig2}e and f. They demonstrate qualitatively that disorder in monolayer TMDs can be visualized directly with a rather simple and efficient imaging technique. A quantitative analysis of the spatial distribution of $P_l$ can be carried out to yield maps analogous to Fig.~\ref{fig2}c and d (see the Supplementary Information).

It is also straight forward to apply the technique in the circular basis. The images of Fig.~\ref{fig4}c and d show the PL intensity of $A$ excitons in $\sigma^-$ co-circular and cross-circular polarimetry for a magnetic field of $+9$~T. The normalized difference of the images yields the $P_c$ profile equivalent to Fig.~\ref{fig3}b (see the Supplementary Information). However, the non-thermalized $A$ exciton valley population of the upper Zeeman branch $K'$ addressed by $\sigma^-$ excitation can be readily deduced from Fig.~\ref{fig4}c and d; for a fully thermalized valley population the flake in Fig.~\ref{fig4}c would be less intense than in Fig.~\ref{fig4}d. The analogous set of measurements for the $L$ exciton band in $\sigma^+$ co- and cross-polarized configurations (Fig.~\ref{fig4}e and f) illustrates the negative degree of circular polarization: the cross-polarized image of the flake is brighter (except for the puddle and a few hot-spots of point-like quantum dot emission) in accord with the raster-scan polarization image of Fig.~\ref{fig3}c.

\vspace{13pt} {\bf \large{Conclusions}}

Our study identifies both raster-scan and wide-field polarimetric imaging as viable tools to explore the valley pseudospin physics in layered TMD semiconductors. Experiments that do not require the full spectral information but are meaningful within a limited PL bandwidth will benefit from a decrease of the measurement time required to achieve the same signal-to-noise performance in wide-field and hyperspectral raster-scan polarimetry. For $(20 \times 20)~\mu$m$^2$ sample areas like in Fig.~\ref{fig4} we estimate a speed-up by at least three orders of magnitude for wide-field imaging as compared to raster-scanning (see the Supplementary Information) which enables rapid large-scale monitoring of the sample (see the Supplementary Video). Clearly, our polarimetric studies are not limited to MoS$_2$ crystals but establish novel analytic means for the entire class of layered TMD semiconductors and heterostructures, and can be extended to spin-polarization imaging of conventional semiconductor quantum wells \cite{Meier1984}. It will perform most efficiently when applied on TMD materials with spectrally separated exciton and trion emission bands, a condition currently accessible with MoS$_2$ only by chemical treatment \cite{Cadiz2016}. By virtue of simplicity and efficiency our technique will facilitate the analysis of TMD materials in terms of crystalline and environmental disorder and, combined with a microscopic model for the optical valley initialization, it will enable quantitative access to site-dependent valley pseudospin dynamics. Moreover, our findings encourage the experimental exploration of topological exciton-polaritons by identifying sufficiently large regions of low valley depolarization and decoherence in monolayer TMDs for the realization of periodic exciton potential arrays that are key to the implementation of topolaritonic devices \cite{Karzig2015}.



\clearpage

{\bf Acknowledgments:} We thank P.~M.~Ajayan for support in the establishment of materials synthesis conditions used in this study, P.~Altpeter and R.~Rath for assistance in the clean room, J.~P.~Kotthaus, B.~Urbaszek and F.~Wang for useful discussions, and P.~Maletinsky and K.~Karrai for valuable input on the manuscript. We gratefully acknowledge funding by the European Research Council under the ERC Grant Agreement no. 336749, the Volks\-wa\-gen Foundation, the DFG Cluster of Excellence Nanosystems Initiative Munich (NIM), and financial support from the Center for NanoScience (CeNS) and LMUinnovativ.

\vspace*{\stretch{1}}

{\bf Author contributions:} A.~N. and A.~H. conceived the experiments. A.~N. built the experimental setup. H.~Y. organized the material aspect and prepared MoS$_2$ flakes on SiO$_2$/Si substrates with support from A.~D.~M.. S.~N. and J.~Lou provided inputs on growth parameters of MoS$_2$ flakes at an initial stage of the project. A.~N., M.~N., and H.~Y. performed basic characterization of the sample. A.~N., J.~Lin., and L.~C. performed the measurements. A.~N., J.~Lin., L.~C., and A.~H. analyzed the data. A.~N. and A.~H. prepared the figures and wrote the manuscript. All authors commented on the manuscript.

\vspace*{\stretch{1}}

{\bf Data availability statement:} The data that support the plots within this paper and other findings of this study are available from the corresponding authors upon reasonable request.

\vspace*{\stretch{1}}

{\bf Additional information:} Supplementary information is available in the online version of the paper. Reprints and permission information is available online at www.nature.com/reprints. Correspondence and requests for materials should be addressed to H.~Y. (hyamaguchi@lanl.gov) and A.~H. (alexander.hoegele@lmu.de).

\vspace*{\stretch{1}}

\clearpage

\clearpage

\clearpage

\begin{figure}[t]
\begin{center}
\includegraphics[scale=1.0]{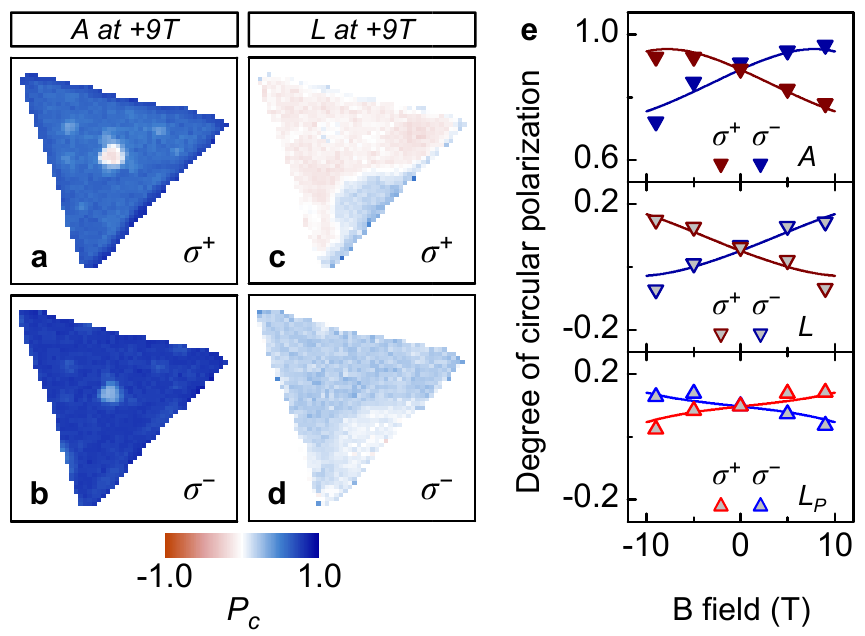}
\caption{\textbf{The valley Zeeman effect in polarimetric imaging.} \textbf{a}, \textbf{b}, and \textbf{c}, \textbf{d}, Circular polarimetric profiles within the $A$ and $L$ exciton bands under $\sigma^+$ and $\sigma^-$ excitations in a magnetic field of $+9$~T. Note the negative $P_c$ of bilayer and $L$ excitons away from the puddle (red-colored regions). \textbf{e}, Evolutions of $A$ and $L$ exciton $P_c$ with magnetic field. Upper and central panels: $A$ and $L$ excitons in a monolayer region away from the puddle; lower panel: $L_P$ excitons in the puddle (the contribution of trions was removed from the $P_c$ of $A$). The solid lines are results of the model as described in the text. All data were recorded at $3.1$~K with excitation at $637$~nm.}
\label{fig3}
\end{center}
\end{figure}

\clearpage


\begin{figure}[t]
\begin{center}
\includegraphics[scale=1.0]{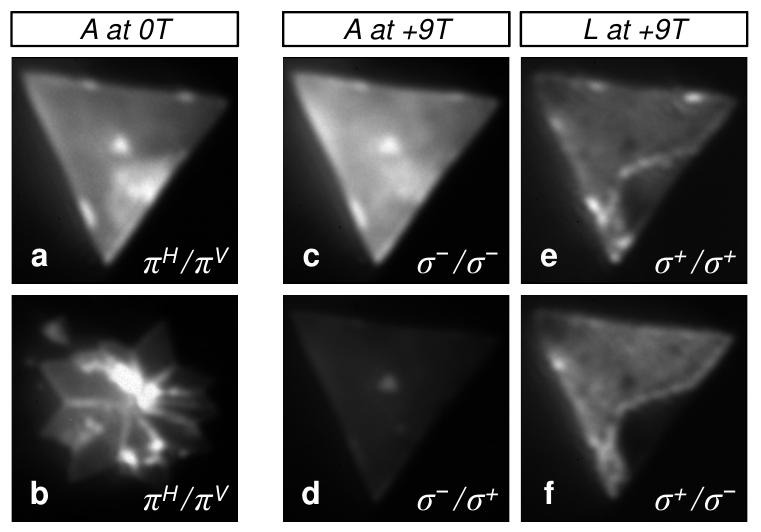}
\caption{\textbf{Wide-field linear and circular polarimetry.} \textbf{a}, \textbf{b}, Images of the $A$ exciton band in cross-linear ($\pi^H$ excitation / $\pi^V$ detection) configuration for the triangular and the star-shaped crystal, respectively. The bright regions visualize bilayer and valley decoherence sites as in Fig.~\ref{fig2}e and f. \textbf{c}, \textbf{d}, Images of the $A$ exciton band in the single-crystal triangle recorded in co-circular ($\sigma^-$ excitation / $\sigma^-$ detection) and cross-circular ($\sigma^-$ excitation / $\sigma^+$ detection) configurations in a magnetic field of $+9$~T. \textbf{e}, \textbf{f}, Same for the $L$ exciton band in $\sigma^+ / \sigma^+$ and $\sigma^+ / \sigma^-$ configurations, respectively. For all images the sample was cooled to $3.1$~K; laser excitation was at $637$~nm.} \label{fig4}
\end{center}
\end{figure}


\clearpage

{\bf Methods:} Monolayer MoS$_2$ crystals on SiO$_2$/Si substrates were prepared by means of CVD and subsequent transfer with polymethyl methacrylate (PMMA). Briefly, SiO$_2$/Si substrates with MoO$_3$ seeding particles were placed in a quartz furnace in the presence of sulfur powder. The furnace was heated to $\sim 900$~$^\circ$C for $15$~min with a flow of inert gases (N$_2$ and Ar) under atmospheric pressure. After cooling down to room temperature, MoS$_2$ crystals were transferred onto p-doped SiO$_2$/Si substrates with a conventional PMMA transfer method. The PMMA was removed by rinsing the sample with MoS$_2$ on SiO$_2$/Si in an acetone bath for three cycles of $15$~min.

Cryogenic confocal spectroscopy, raster-scan and wide-field imaging were performed in an ultra-low vibration closed-cycle cryostat (attocube systems, attoDRY1000) with a base temperature of $3.1$~K and a superconducting magnet with fields of up to $\pm 9$~T. The sample was positioned with nanopositioners (attocube systems, ANP101 series) into the focal plane of a low-temperature apochromat with a numerical aperture of $0.65$ (attocube systems, LT-APO/VIS/0.65) and confocal excitation and detection spots of $0.7~\mu$m full-width at half-maximum diameters. Linear polarizers (Thorlabs, LPVIS and LPVISB), half- and quarter-waveplates (B.~Halle, RAC~3~series) mounted on piezo-rotators (attocube systems, ANR240) were used to control the photon polarization in the excitation and detection pathways. PL spectroscopy was performed with continuous wave excitation lasers at $532$~nm (CNI, \mbox{MLL-III-532-50-1}) or $637$~nm (New Focus, Velocity TLB-6704), and a standard spectrometer (PI, Acton SP-2558) with a  nitrogen cooled silicon CCD (PI, Spec-10:100BR/LN). The spectral resolution of the system was $0.35$~meV.

For wide-field polarimetric imaging, a femtosecond optical parametric oscillator operated at $637$~nm (Coherent, Mira-OPO) was defocussed to illuminate a sample area of $\sim 4.4 \cdot 10^{-3}$~mm$^2$ at an average excitation power density of $\sim 200$~W/cm$^2$. The PL was imaged with $50\times$ magnification onto a silicon CCD (Point Grey, GS3-U3-14S5M-C) within variable spectral bands set by tunable band-pass filters (Semrock, VersaChrome). The images of Fig.~\ref{fig4} were recorded with an encoding gamma of $0.7$. The intensities were scaled to the full dynamic range of the grayscale by $1.05$ for Fig.~\ref{fig4}a, c, and d, and $1.54$ for Fig.~\ref{fig4}e and f, respectively. The image in Fig.~\ref{fig4}b was not scaled.


\renewcommand{\figurename}{Supplementary Fig.}
\newcommand{\figref}[1]{Supplementary Fig.~\ref{#1}}

\begin{supplement}{10}

\begin{center}
\vspace*{20pt}
{\LARGE SUPPLEMENTARY INFORMATION \\ \vspace{36pt} Opto-valleytronic imaging of atomically thin semiconductors}

\vspace{14pt}
{\large Andre Neumann, Jessica Lindlau, L\'{e}o Colombier, Manuel Nutz,\\ \vspace{4pt} Sina Najmaei, Jun Lou, Aditya D. Mohite, Hisato Yamaguchi, and Alexander H\"{o}gele}
\end{center}

\vspace{12pt}
\section{Sample fabrication and basic characterization}\label{sample}

The chemical vapor deposition (CVD) of monolayer MoS$_2$ crystals was adopted from Najmaei and coworkers \cite{SI_Najmaei2013}. Briefly, SiO$_2$/Si substrates with MoO$_3$ seeding particles were placed in a quartz furnace in the presence of sulfur powder. The furnace was heated to $\sim 900$~$^\circ$C for $15$~min with a flow of inert gases (N$_2$ and Ar) under atmospheric pressure. After cooling down to room temperature, synthesized MoS$_2$ crystals were transferred onto p-doped SiO$_2$/Si substrates with a conventional polymethyl methacrylate (PMMA) transfer method \cite{SI_Reina2008}. The PMMA was removed by rinsing the resulting sample with MoS$_2$ on SiO$_2$/Si in an acetone bath for three cycles of $15$~min. Optical and atomic force microscopy (AFM) images of typical MoS$_2$ monolayer flakes after transfer onto SiO$_2$/Si are shown in \figref{sfig1}.

\begin{figure}[!h]
\begin{center}
\vspace{10pt}
\includegraphics[scale=1.15]{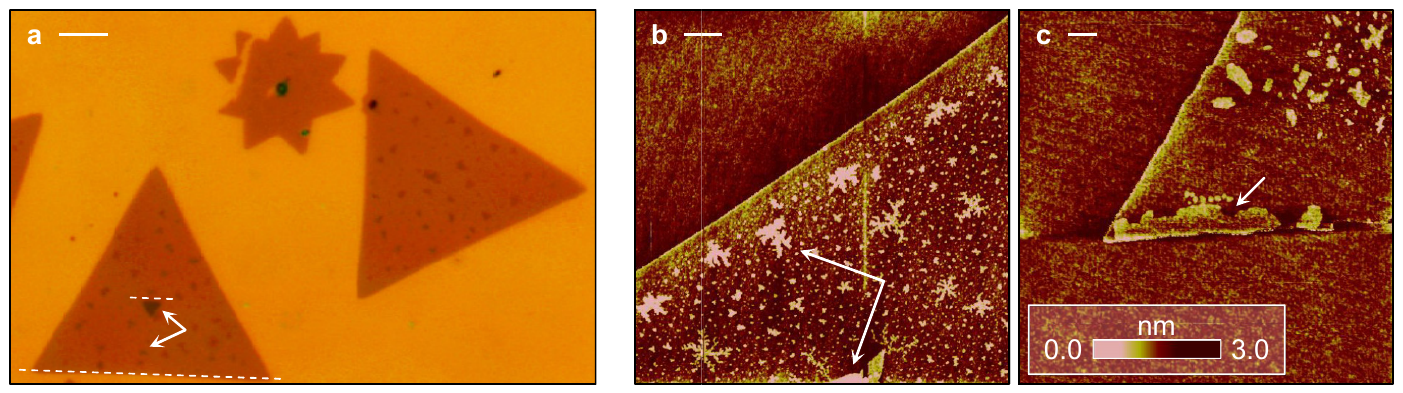}
\vspace{-10pt} \caption{\textbf{a}, Optical image and \textbf{b}, \textbf{c}, atomic force micrographs of monolayer MoS$_2$ grown by chemical vapor deposition. The scale bars are $10~\mu$m in \textbf{a}, and $2~\mu$m in \textbf{b} and \textbf{c}. Arrows indicate seeds of bilayer formation on top of monolayer flakes in \textbf{a} and \textbf{b}. Note the $180^{\circ}$ misorientation of the large monolayer and the small bilayer triangles characteristic of ideal Bernal stacking indicated by the white dashed lines in \textbf{a}. An example of an extended region of surface contamination of a single-crystal triangle is indicated by the arrow at the at the lower edge in \textbf{c}.}
\label{sfig1}
\end{center}
\end{figure}

\section{Experimental setup and settings}\label{setup}

Cryogenic confocal spectroscopy, raster-scan and wide-field imaging were performed in an ultra-low vibration closed-cycle cryostat (attocube systems, attoDRY1000) with a base temperature of $3.1$~K and a superconducting magnet with fields of up to $\pm 9$~T. The sample was positioned with nanopositioners (attocube systems, ANP101 series) into the focal plane of a low-temperature apochromat with a numerical aperture of $0.65$ (attocube systems, LT-APO/VIS/0.65) and confocal excitation and detection spots with $0.7~\mu$m full-width at half-maximum (FWHM) diameters. Linear polarizers (Thorlabs, LPVIS and LPVISB), half- and quarter-waveplates (B.~Halle, RAC~3~series) mounted on piezo-rotators (attocube systems, ANR240) were used to control the photon polarization in the excitation and detection pathways. PL spectroscopy was performed with continuous wave (cw) excitation lasers at $532$~nm (CNI, \mbox{MLL-III-532-50-1}) or $637$~nm (New Focus, Velocity TLB-6704), and a standard spectrometer (PI, Acton SP-2558) with a liquid nitrogen cooled silicon CCD (PI, Spec-10:100BR/LN, gain setting of $4$~e$^-$/count). The spectral resolution of the system was $\sim 0.35$~meV for confocal PL and $\sim 0.6$~cm$^{-1}$ for Raman measurements. A streak camera (Hamamatsu, C5680-24S) and a femtosecond (fs) optical parametric oscillator (OPO) operated at $630$~nm or $637$~nm (Coherent, Mira-OPO) were used for time-resolved PL studies. The fiber-coupled system exhibited dispersion-limited temporal resolution down to $\sim 10$~ps depending on the spectral bandwidth and the length of the single-mode fiber. For wide-field imaging, the OPO was tuned to $637$~nm and defocussed to illuminate a sample area of $\sim 4.4 \cdot 10^{-3}$~mm$^2$ at an average excitation power density of $\sim 200$~W/cm$^2$. PL images with $50\times$ magnification were acquired in the spectral bands of tunable band-pass filters (Semrock, VersaChrome) with a silicon CCD (Point Grey, GS3-U3-14S5M-C, pixel size of $\left( 6.45 \times 6.45 \right)~\mu$m$^2$ and saturation capacity of $17$~ke$^-$). All wide-field images were recorded with an integration time of $500$~s ($300$~s in \figref{sfig10}b) and a CCD gain setting of $0$~dB (except for $5$~dB in Fig.~4a of the main text and $3$~dB in \figref{sfig10}a~-~d). The images digitized to a $16$~bit format with $14$~bit conversion were acquired with linear gamma encoding (gamma encoding of $0.7$ in Fig.~4 of the main text). The intensities were multiplied with scaling factors ranging from $1.0$ to $1.7$ to span the full dynamic range of the grayscale (identical factors were used for co- and cross-polarized images of the same exciton bands).

\section{Spectroscopy and polarimetry experiments}\label{polarimetry}
\vspace{-6pt}
\subsection{Basic spectral characteristics}\label{basic}

The PL of monolayer MoS$_2$ crystals on SiO$_2$/Si was excited either non-resonantly at $532$~nm ($2.33$~eV) or at $637$~nm ($1.95$~eV) in resonance with the blue shoulder of the $A$ exciton. A non-resonant spectrum recorded on a regular position of the triangular flake is shown in \figref{sfig2}a (same as in Fig.~1b of the main text). The non-resonant PL exhibits contributions of both neutral and charged excitons, $A$ and $A^-$ due to unintentional doping of MoS$_2$ crystals on p-doped SiO$_2$/Si substrates \cite{SI_Scheuschner2014}. The asymmetric lineshape of the total PL (spectrum in \figref{sfig2}a) can be decomposed into two Lorentzians with equal FWHM linewidths of $\sim 62$~meV separated by the trion binding energy of $\sim 30$~meV \cite{SI_Berkelbach2013}. The near-resonant spectrum recorded with a long-pass filter at $652$~nm ($1.90$~eV) in \figref{sfig2}b (gray and blue traces show for the same flake position non-resonant and near-resonant spectra, respectively) exhibited additional sharp features identified as Raman scattered photons by the substrate (Si) and the MoS$_2$ monolayer, with assignments given in \figref{sfig2}c adopted from Ref.~\citenum{SI_Golasa2014}. The Raman spectra were useful to identify bilayer regions (as in the center of the single-crystal triangle) where they exhibited a characteristic splitting of the $A_{1g}(\Gamma)$ mode shown in \figref{sfig2}d under resonant excitation with the fundamental exciton \cite{SI_Staiger2015}. To avoid contamination of the opto-valleytronic properties of $A$ excitons by Raman photons, we selected a spectral band of $3$~meV width centered at $1.879$~eV near the PL maximum of the $A$ exciton emission and away from Raman resonances (the red spectral interval in \figref{sfig2}a~-~c is the same as in Fig.~1b of the main text ). Moreover, for the quantitative analysis of the degree of circular polarization of $A$ excitons presented in Fig.~3e of the main text, the contribution of $A^-$ was removed by fitting the total PL with two Lorentzians and subtracting the contribution of the trion PL from the spectral band chosen for the opto-valleytronic analysis as described above.

\begin{figure}[!b]
\begin{center}
\includegraphics[scale=1.15]{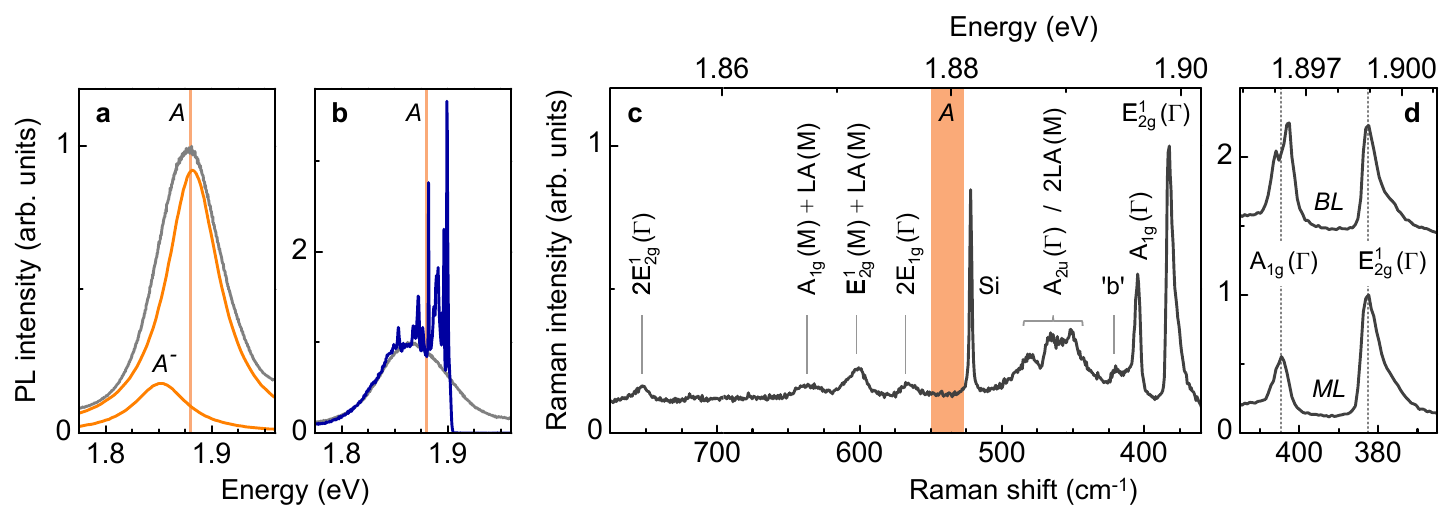}
\vspace{-10pt} \caption{\textbf{a}, Photoluminescence spectrum reproduced from Fig.~1 of the main text (gray) with Lorentzian fits (orange) to the contributions of the neutral and charged excitons, $A$ and $A^-$, centered around $1.882$~eV and $1.852$~eV, respectively, with full-width at half-maximum linewidths of $62$~meV. \textbf{b}, Comparison of the photoluminescence spectra for non-resonant excitation at $532$~nm ($2.33$~eV) and excitation at $637$~nm ($1.95$~eV) in resonance with the blue shoulder of the $A$ exciton (gray and blue traces, respectively); the near-resonant spectrum was recorded with a long-pass filter at $652$~nm ($1.90$~eV). \textbf{c}, Raman spectrum for the near-resonant excitation at $637$~nm is superimposed as sharp spectral features on the photoluminescence peak in \textbf{b}. The Raman modes were assigned according to the MoS$_2$ bulk notation of Ref.~\citenum{SI_Golasa2014}. The data in \textbf{b} and \textbf{c} were measured on the same regular position of the monolayer triangle from the main text. A spectral band of $3$~meV width (indicated in red) was selected at the maximum emission of the $A$ exciton and away from Raman features to construct the polarimetric maps without contamination by Raman photons. \textbf{d}, Vertically offset Raman spectra of the $E^{1}_{2g}(\Gamma)$ and $A_{1g}(\Gamma)$ modes for the near-resonant excitation at $637$~nm on representative monolayer (ML) and bilayer (BL) positions of the triangular flake (the dashed lines are guides to the eye). The splitting of the $A_{1g}(\Gamma)$ mode is characteristic of bilayer transition metal dichalcogenides under resonant excitation \cite{SI_Staiger2015}. All measurements were obtained at $3.1$~K.}
\label{sfig2}
\end{center}
\end{figure}

\subsection{Time-resolved photoluminescence measurements}\label{timeresolved}

The PL decay dynamics of the monolayer MoS$_2$ triangle from the main text are presented in \figref{sfig3}. The $A$ exciton PL, measured in a spectral band of $35$~meV, exhibited radiative decay with a time constant below the resolution limit of $\sim 14$~ps given by the dispersion-limited instrument response function (\figref{sfig3}a). Localized excitons, selected by a bandpass filter of $50$~meV width, exhibited two decay timescales (\figref{sfig3}b): the dominant fast component was resolution-limited (with partial contribution from $A$ excitons), and the slow component exhibited monoexponential decay with a lifetime of $174 \pm 12$~ps in agreement with previous results \cite{SI_Lagarde2014}. The slow decay channel accounted for $8$\% of the total PL intensity on a regular position ($L$) and for $12$\% in the puddle ($L_P$) of the triangular crystal.

\begin{figure}[!h]
\begin{center}
\vspace{10pt}
\includegraphics[scale=1.15]{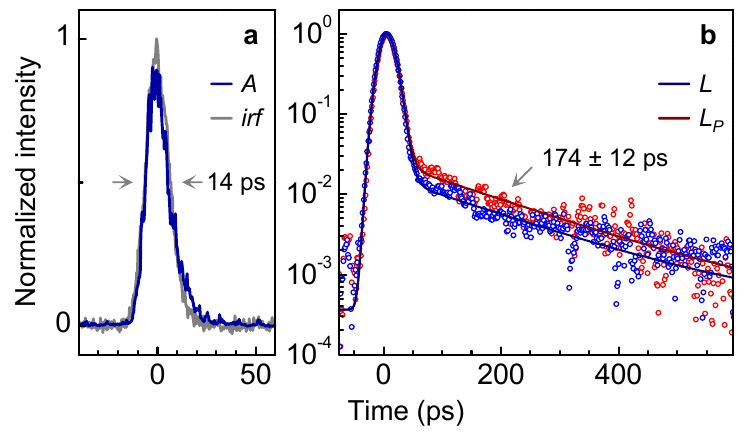}
\vspace{-10pt} \caption{\textbf{a}, \textbf{b}, Time-resolved photoluminescence of $A$ excitons (blue trace) and localized excitons in and away from the puddle, $L_P$ and $L$ (red and blue circles). Band-pass filters with widths of $35$~meV ($50$~meV) were used for spectral selection of $A$ ($L$ and $L_P$) excitons. The decay dynamics in \textbf{a} were limited by the instrument response function (irf) with a measured full-width at half-maximum of $\sim 14$~ps (gray trace). The solid lines in \textbf{b} are fits to the data: the fast decay components of the localized excitons were dispersion broadened, the slow components exhibited monoexponential decays with decay constants of $174 \pm 12$~ps and contributions relative to the total signal of $12$\% ($L_P$, dark red trace) and $8$\% ($L$, dark blue trace). All data were measured on the triangular MoS$_2$ flake of the main text with laser excitation at $637$~nm in \textbf{a} and $630$~nm in \textbf{b}; the temperature was $3.1$~K.}
\label{sfig3}
\end{center}
\end{figure}

\subsection{Spectral characteristics of polarization-resolved photoluminescence}\label{spectral}

\figref{sfig4}a and c show co-polarized ($I_{co}$) and cross-polarized ($I_{cr}$) PL spectra recorded with an excitation laser at $637$~nm ($1.95$~eV) in circular and linear bases, respectively. The spectral characteristics of $P_{c}$ and $P_{l}$, respectively shown in \figref{sfig4}b and d, were calculated as the normalized differences between co- and cross-polarized PL intensities according to $P=(I_{co} - I_{cr})/(I_{co} + I_{cr})$. Data in \figref{sfig4}e and f confirm that the degree of linear polarization is independent of the choice of the linear basis (the linear PL polarization is parallel to the axis of the excitation laser set along $\pi^H$ and $\pi^D$ in \figref{sfig4}e and f, respectively). The red bar in \figref{sfig4}a~-~d was used for confocal opto-valleytronic imaging of $A$ excitons (same band as in \figref{sfig2}a~-~c and Fig.~1b of the main text).

\begin{figure}[!h]
\begin{center}
\vspace{10pt}
\includegraphics[scale=1.15]{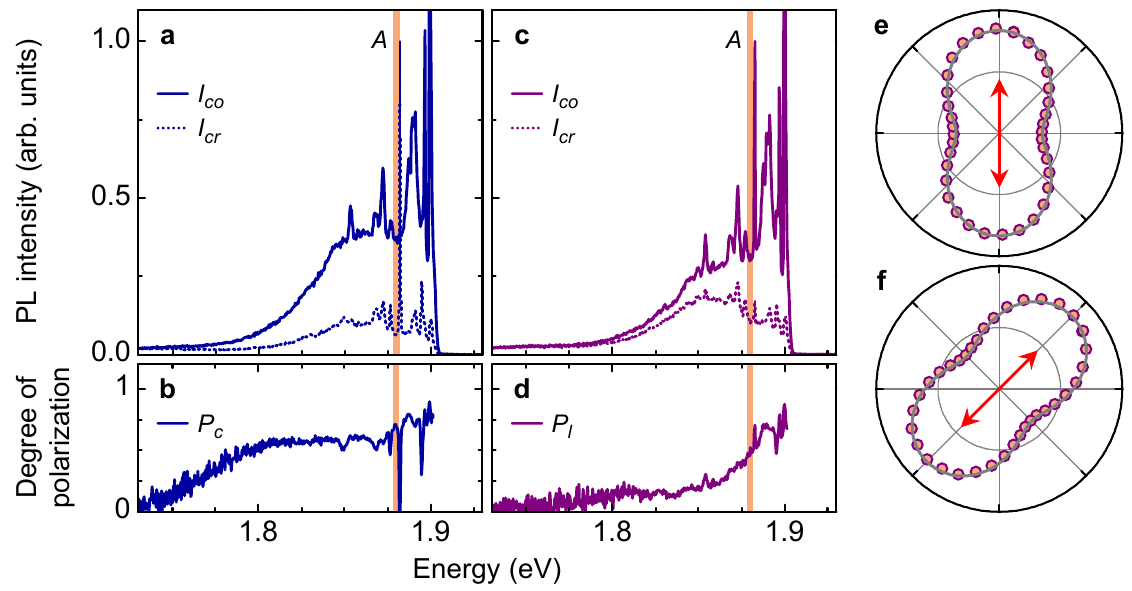}
\vspace{-10pt} \caption{\textbf{a}, Photoluminescence spectra recorded in co-polarized (solid trace, $I_{co}$) and cross-polarized (dashed trace, $I_{cr}$) configurations with $\sigma^+$ excitation. \textbf{b}, Corresponding degree of circular polarization $P_{c}$. \textbf{c}, Co- and cross-polarized photoluminescence spectra under linear ($\pi^H$) excitation and \textbf{d}, degree of linear polarization $P_{l}$. The bands of $3$~meV width indicated in red (same as in \figref{sfig2} and Fig.~1 of the main text) were used to select $A$ excitons for confocal opto-valleytronic imaging. \textbf{e}, \textbf{f}, Polar plots of the normalized PL intensity within the spectral band as a function of the rotation angle $\theta$ of the linear analyzer for $\pi^H$, $\pi^D$ orientations of the linear polarizer (indicated by red arrows), respectively. The gray solid lines are fits to the data with a $\left[ 1 + P_{l} \cdot \cos \left( 2 \theta - 2 \phi \right) \right]$ functional dependence, where $\phi$ is the polarizer angle. All measurements were recorded on the monolayer triangle away from defects as discussed in the main text with an excitation laser at $637$~nm ($1.95$~eV) and a long-pass filter at $652$~nm ($1.90$~eV); the temperature was $3.1$~K.}
\label{sfig4}
\end{center}
\end{figure}

\subsection{Circular and linear raster-scan polarimetry}\label{confocal}

To construct polarimetric maps we raster-scanned the sample with respect to fixed confocal excitation and detection spots, and performed spectral acquisition of co- and cross-polarized PL at each raster pixel with an excitation power density of $\sim 2.5 \cdot 10^4$~W/cm$^2$. An averaged background spectrum was subtracted from all image pixels, and co- and cross-polarized PL intensities of $A$ and $L$ excitons were integrated within the spectral bands shown in Fig.~1b of the main text. Pixels with a standard deviation of $P$ above 0.05 (stemming from vanishingly small PL intensities away from the flake) were set to zero. In \figref{sfig5}a~-~d we reproduce the maps of Fig.~2a~-~d of the main text to indicate specific positions (numbered from $1$ to $6$) for a quantitative comparison of site-to-site variations of $P_c$ (blue bars) and $P_l$ (purple bars) summarized in \figref{sfig5}e for the of $A$ exciton band. Similar polarimetric measurements were performed on other MoS$_2$ flakes. \figref{sfig6}a,~c and \figref{sfig6}b,~d show profiles of $P_c$ and $P_l$ for a poly-crystalline and a single-crystalline MoS$_2$ monolayer, respectively. The bar chart in \figref{sfig6}e summarizes the statistics of $P_c$ and $P_l$ acquired on five different monolayer flakes.

\begin{figure}[!ht]
\begin{center}
\includegraphics[scale=1.15]{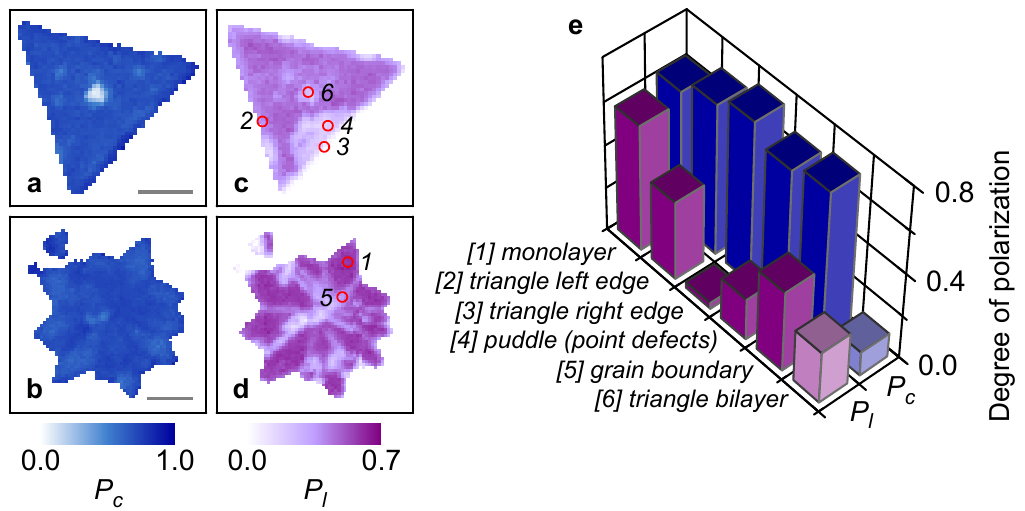}
\vspace{-10pt} \caption{\textbf{a}, \textbf{b}, Circular and \textbf{c}, \textbf{d}, linear polarimetric profiles for $A$ exciton bands of single- and poly-crystalline MoS$_2$, respectively. The data are reproduced from Fig.~2 of the main text with specific positions $1 - 6$ added for a quantitative comparison (scale bars are $5~\mu$m). \textbf{e}, $P_{c}$ (blue bars) and $P_{l}$ (purple bars) for $A$ excitons at characteristic positions of the flakes marked with red circles in \textbf{c} and \textbf{d}. All data were measured at $3.1$~K with laser excitation at $637$~nm.}
\label{sfig5}
\end{center}
\end{figure}

\begin{figure}[!ht]
\begin{center}
\includegraphics[scale=1.15]{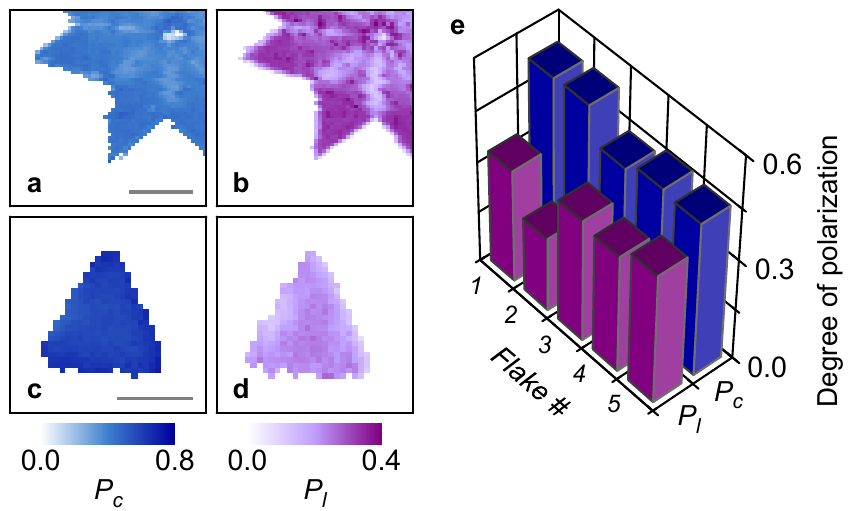}
\vspace{-10pt} \caption{\textbf{a}, \textbf{b}, Maps of $P_{c}$ and $P_{l}$ for a segment of a poly-crystalline MoS$_2$ flake (scale bar is $10~\mu$m). \textbf{c}, \textbf{d}, Maps of $P_{c}$ and $P_{l}$ for a single-crystal monolayer triangle (scale bar is $3~\mu$m). The bar chart in \textbf{e} summarizes variations in $P_{c}$ and $P_{l}$ (blue and purple bars, respectively) for five different MoS$_2$ monolayers at $3.1$~K and $637$~nm excitation.} \label{sfig6}
\end{center}
\end{figure}

\subsection{Equivalence of polarization bases at zero magnetic field}\label{dop_nofield}

In the absence of an external magnetic field, time-reversal symmetry implies the equivalence of the degrees of circular polarization measured in $\sigma^+$ and $\sigma^-$ configurations. The set of data in \figref{sfig7}a~-~c demonstrates that the degree of the circular polarization is independent of the choice of the circular basis: $P_{c}$ is identical (within the precision of our measurement) for $\sigma^+$ and $\sigma^-$ polarization bases (compare \figref{sfig7}a and b, respectively). This fact is also reflected by the vanishing difference profile $\Delta P_{c} = P_{c}(\sigma^+) - P_{c}(\sigma^-)$ shown in \figref{sfig7}c. The analogous set of data is respectively shown for the degrees of linear polarization recorded under horizontal ($\pi^H$) and diagonal ($\pi^D$) linearly polarized excitations in \figref{sfig7}d and e; their vanishing difference $\Delta P_{l} = P_{l} (\pi^H) - P_{l} (\pi^D)$ is plotted in \figref{sfig7}f. The data demonstrate the independence of $P_l$ of the choice of the linear basis.

\begin{figure}[!h]
\begin{center}
\vspace{10pt}
\includegraphics[scale=1.15]{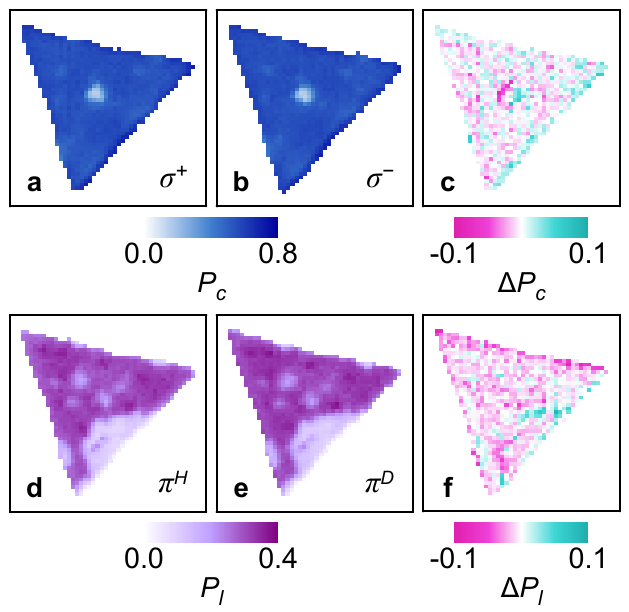}
\vspace{-10pt} \caption{\textbf{a}, \textbf{b}, $P_{c}$ and \textbf{d}, \textbf{e}, $P_{l}$ for circularly polarized $\sigma^+$, $\sigma^-$ (right- and left-handed) and linearly polarized $\pi^H$, $\pi^D$ (horizontal and diagonal) laser excitation. \textbf{c}, Equivalence of the circular polarization bases, where changes to $P_{c}$ are prohibited by time-reversal symmetry, computed as $\Delta P_{c} = P_{c}(\sigma^+) - P_{c}(\sigma^-) \simeq 0$. \textbf{f}, Same for different linear polarization bases: $\Delta P_{l} = P_{l} (\pi^H) - P_{l} (\pi^D) \simeq 0$. Note the changed scale in \textbf{c} and \textbf{f}. All measurements were spectrally integrated and recorded at zero magnetic field with laser excitation at $637$~nm. The temperature was $3.1$~K.}
\label{sfig7}
\end{center}
\end{figure}

\subsection{Magnetic field dependence of the degree of circular polarization}\label{dop_withfield}

A finite external magnetic field applied in Faraday geometry (perpendicular to the TMD monolayer surface) lifts the valley degeneracy and gives rise to a Zeeman shift of opposite sign for excitons in the $K$ and $K'$ valleys. A positive magnetic field decreases (increases) the energy of the $K$ ($K'$) valley exciton. Consequently, an applied magnetic field changes the degrees of circular polarization of the valley exciton emission. The changes are discussed in the main text for spectrally selected $A$ and $L$ excitons under $\sigma^+$ and $\sigma^-$ excitation at $+9$~T (the corresponding data of Fig.~3 are reproduced in \figref{sfig8}a, b, d, and e). In \figref{sfig8}c and f we plot for both types of excitons the change in the degree of circular polarization, quantified as $\Delta P_{c} = P_{c} ({\sigma^+})-P_{c} ({\sigma^-})$. For $A$ excitons at $B=+9$~T, $\Delta P_{c}$ is negative (magenta-colored) throughout the MoS$_2$ flake (\figref{sfig8}c). This corresponds to a decrease (increase) of the $\sigma^+$ ($\sigma^-$) degree of circular polarization for the lower (upper) $K$ ($K'$) Zeeman branch of $A$ excitons. The $L$ excitons show the same trend away from the puddle and reversed features in the puddle (cyan-colored region in \figref{sfig8}f) in accord with thermal population distribution discussed in the main text. The right panel of \figref{sfig8} shows the same set of data but for $B=-9$~T. The sign reversal of the magnetic field results in interchanged roles of $\sigma^+$ and $\sigma^-$ polarizations (\figref{sfig8}g, h, j, and k) and thus in a sign reversal of $\Delta P_{c}$ (\figref{sfig8}i and l).

\begin{figure}[!h]
\begin{center}
\vspace{10pt}
\includegraphics[scale=1.15]{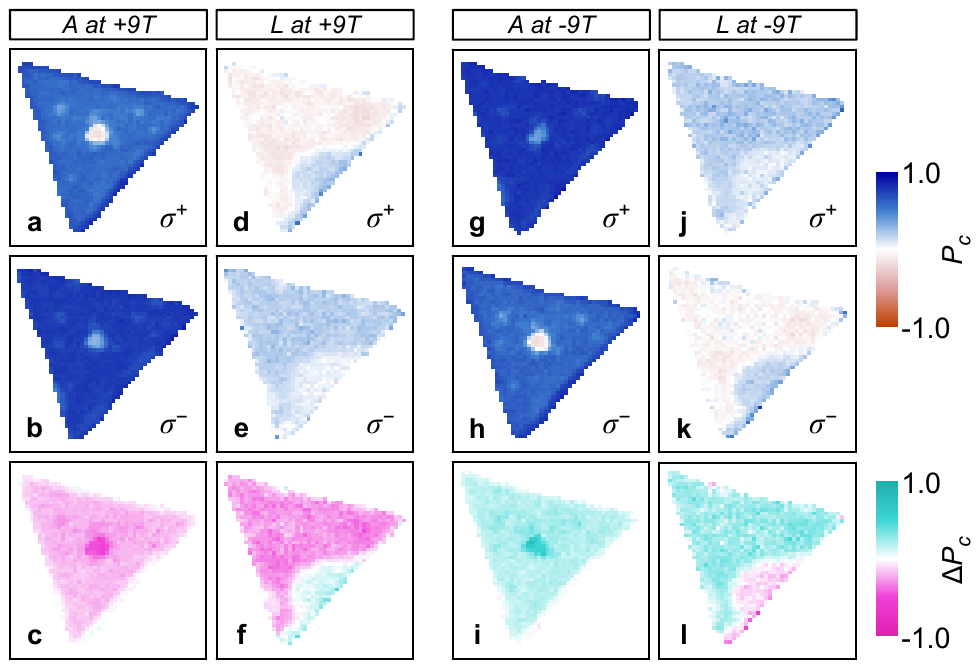}
\vspace{-10pt} \caption{Left panel: \textbf{a}, \textbf{b}, and \textbf{d}, \textbf{e}, Circular polarimetric profiles within the $A$ and $L$ exciton bands under $\sigma^+$ and $\sigma^-$ excitations in a magnetic field of $B=+9$~T reproduced from Fig.~3 of the main text. \textbf{c}, \textbf{f}, Corresponding changes in the degrees of circular polarization of the photoluminescence within the $A$ and $L$ exciton bands, respectively, computed as $\Delta P_{c} = P_{c} (\sigma^+) - P_{c} (\sigma^-)$. Right panel: \textbf{g}~-~\textbf{l}, Same as the left panel \textbf{a}~-~\textbf{f}, but for $B=-9$~T. Sign reversal of magnetic field interchanges the roles of the Zeeman branches associated with $\sigma^+$ and $\sigma^-$ excitations, which results in a sign reversal of $\Delta P_{c}$ as compared to the left panel. All data were recorded at $3.1$~K and with a $637$~nm laser.}
\label{sfig8}
\end{center}
\end{figure}

\vspace{-10pt}
\subsection{Wide-field imaging and polarimetry}\label{imaging}

As described previously the confocal setup was modified for wide-field imaging by defocusing the excitation laser to illuminate a sample area of $\sim 4.4 \cdot 10^{-3}$~mm$^2$ and by replacing the single-mode fiber in the detection path with an imaging lens and a silicon CCD. A tunable band-pass filter (with $15$~nm FWHM bandwidth) was inserted before the imaging lens to select exciton-specific bands. The corresponding setup schematics are shown in \figref{sfig9}a. A fs-OPO operated at $637$~nm was used to excite the PL at an average power density of $\sim 200$~W/cm$^2$. This average power density was a factor of $\sim 150$ and $\sim 500$ lower than in confocal measurements with the excitation laser at $637$~nm (cw power density of $\sim 2.5 \cdot 10^4$~W/cm$^2$) and $532$~nm (cw power density of $\sim 9.4 \cdot 10^4$~W/cm$^2$), respectively. In the following, we discuss the main implications of the reduced power density for the observations in wide-field imaging and polarimetry.

\begin{figure}[!b]
\begin{center}
\includegraphics[scale=1.15]{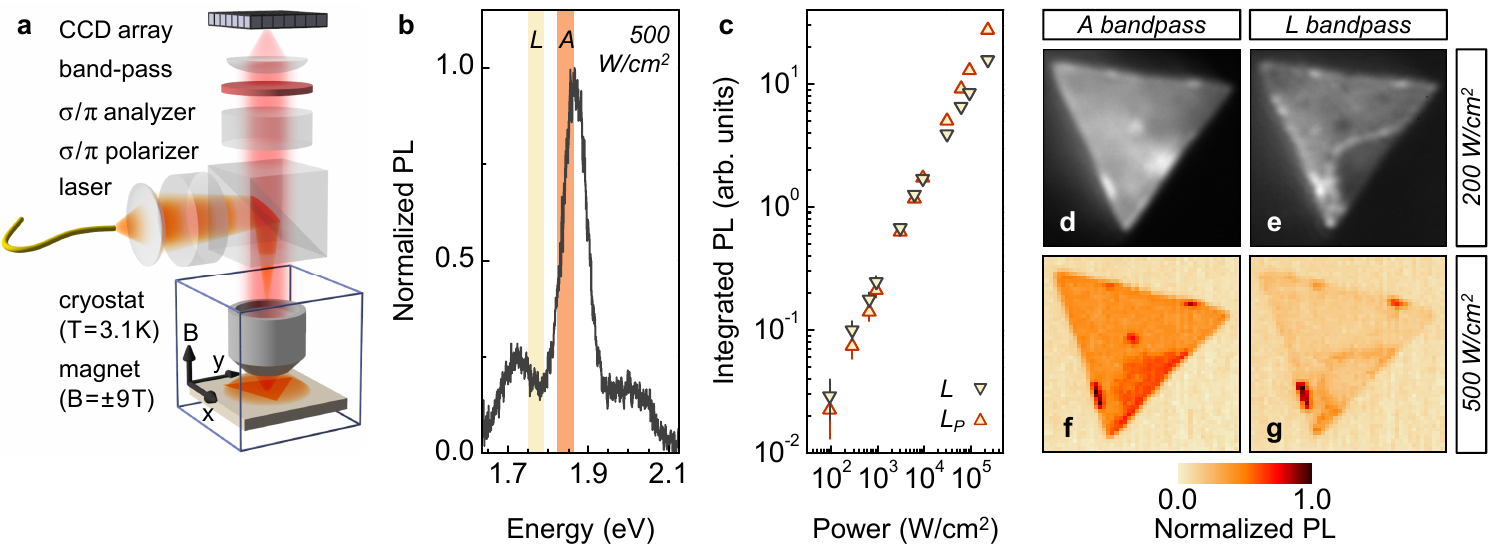}
\vspace{-26pt} \caption{Confocal and wide-field imaging at low excitation power densities of $\sim 500$~W/cm$^2$ and $\sim 200$~W/cm$^2$, respectively. \textbf{a}, Setup schematics for wide-field imaging: the excitation laser was defocussed to illuminate a spot of $\sim 100~\mu$m diameter. Exciton-specific photoluminescence was spectrally selected with a band-pass filter and imaged onto a CCD camera with a $50 \times$ effective magnification. \textbf{b}, Confocal low-power excitation spectrum with colored $A$ and $L$ exciton bands used for exciton-selective photoluminescence imaging. \textbf{c}, Photoluminescence intensity of $L$ excitons as a function of excitation power away from the puddle ($L$) and in the puddle ($L_P$). \textbf{d}, \textbf{f}, Photoluminescence intensity profiles of $A$ excitons for the triangular flake of the main text under low-power excitation in wide-field and confocal imaging, respectively. \textbf{e}, \textbf{g}, Same but for the $L$ exciton band. The excitation wavelength was $637$~nm in \textbf{d} and \textbf{e}, and $532$~nm in \textbf{b}, \textbf{c}, \textbf{f}, and \textbf{g}. The images in \textbf{d} and \textbf{e} were acquired with linear gamma encoding. All data were recorded at $3.1$~K.}
\label{sfig9}
\end{center}
\end{figure}

To compare the confocal and wide-field imaging modes we recorded additional PL spectra and intensity maps in the confocal setup configuration at reduced power densities. In \figref{sfig9}b~-~g we present data recorded in the regime of low power excitation for the single-crystal MoS$_2$ triangle discussed in the main text. A confocal PL spectrum recorded at an excitation power density comparable to that of the wide-field imaging mode is shown in \figref{sfig9}b. Colored bands indicate the spectral intervals that were used for the evaluation of confocal PL intensity maps in \figref{sfig9}f and g for $A$ and $L$ excitons, respectively. The bands were selected to match the setting of the band-pass filter in wide-field imaging.

At low excitation powers the PL intensity of $L$ excitons is not saturated (\figref{sfig9}c). This results in an increased intensity ratio of $L$ to $A$ exciton PL in the confocal spectrum of \figref{sfig9}b as compared to the data in Fig.~1b of the main text. Moreover, since the saturation responses of $L$ excitons away from the puddle and in the puddle are different (the respective data are denoted as $L$ and $L_P$ in \figref{sfig9}c), one expects a crossover in the relative PL intensities of defect-bound excitons in the puddle and on a regular position of the flake. This effect is observed in PL imaging at low excitation power densities: both wide-field (\figref{sfig9}e) and confocal (\figref{sfig9}g) images of the monolayer triangle show a more intense $L$ exciton PL away from the puddle, whereas at high excitation power densities the $L$ exciton PL in the puddle is more intense (Fig.~1e of the main text). The more pronounced appearance of hot-spots in the PL intensity profiles of $A$ excitons under low-power illumination (\figref{sfig9}d and f) as compared to the PL intensity map under high-power illumination (Fig.~1d of the main text) is also attributed to saturation effects.

\begin{figure}[!b]
\begin{center}
\vspace{-10pt}
\includegraphics[scale=1.15]{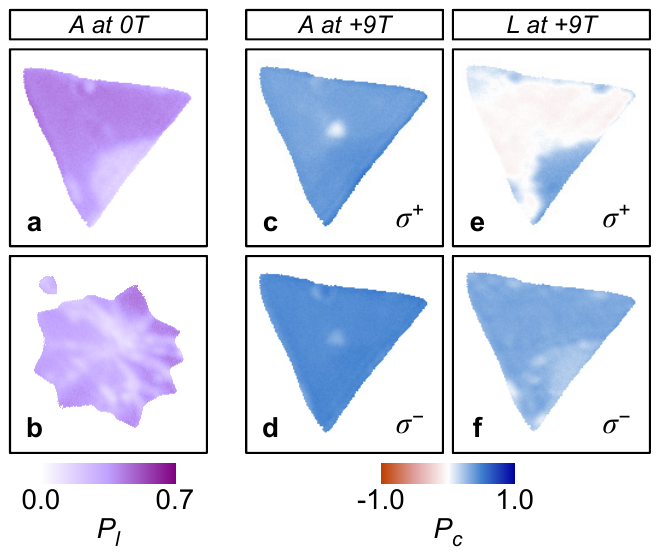}
\vspace{-16pt} \caption{\textbf{a}, \textbf{b}, Linear polarimetric profiles $P_{l}$ of spectrally filtered $A$ excitons in single- and poly-crystalline MoS$_2$, respectively, obtained with wide-field imaging. \textbf{c}, \textbf{d}, and \textbf{e}, \textbf{f}, Circular polarimetric profiles $P_{c}$ of $A$ and $L$ excitons within the respective band-pass windows in an external magnetic field of $+9$~T. The data were recorded at $3.1$~K with excitation at $637$~nm and $\sigma^+$ polarization in \textbf{c} and \textbf{e}, and $\sigma^-$ polarization in \textbf{d} and \textbf{f}.}
\label{sfig10}
\end{center}
\end{figure}

As in confocal polarimetry, polarization-resolved wide-field imaging can be used to construct polarimetric profiles of valley polarization and valley coherence. To demonstrate the quantitative character of the technique, we show in \figref{sfig10}a~-~f the $P_{l}$ and $P_{c}$ profiles of the single-crystal MoS$_2$ triangle of the main text obtained with wide-field polarimetry. Note that the opto-valleytronic profiles are in excellent quantitative agreement with confocal data shown in Fig.~2 and Fig.~3 of the main text if one takes into account the reduction of $P_{l}$ and $P_{c}$ due to wider spectral bands of the band-pass filter used in wide-field polarimetry.

The advantage of the wide-field imaging technique is that it reduces the integration time for opto-valleytronic profiling whenever full spectral information at each pixel of the map is not required. Ultimately, for an image size of $\sim \left( 20 \times 20 \right)~\mu$m$^2$ with a pixel size of $\sim \left( 0.4 \times 0.4 \right)~\mu$m$^2$ an acquisition acceleration of at least $\sim 2.5 \cdot 10^3$ can be obtained with polarimetric profiling in wide-field imaging as compared to confocal raster-scanning with the same spatial resolution, spectral width, noise level, and transmission characteristics. For this estimation we neglected the confocal scanning time and assumed equal excitation power densities of the imaging techniques as well as shot-noise limited signals. We note that for our confocal raster-scan polarimetry the scanning time was a factor of two longer than the integration time, and the transmission properties were a factor of ten lower than for wide-field imaging.

\section{Theoretical modeling}\label{theory}

In the following we model the degree of circular polarization of excitons in monolayer MoS$_2$ in the presence of an external magnetic field applied in Faraday geometry (the magnetic field is oriented perpendicular to the TMD monolayer surface). To this end we consider a four level system with the crystal ground state $\lvert 0 \rangle$, the $K$ exciton state $\lvert 1 \rangle$, the $K'$ exciton state $\lvert 2 \rangle$, and an excited state $\lvert 3 \rangle$ in one of the two valleys. All levels are denoted in \figref{sfig11}, where without loss of generality the excited state $\lvert 3 \rangle$ is in the $K$ valley. In the following, exciton states in $K$ ($K'$) that couple to $\sigma^{+}$ ($\sigma^{-}$) polarized optical transitions are tagged with the valley index $\kappa=+ 1$ ($\kappa=- 1$).

\begin{figure}[!h]
\begin{center}
\includegraphics[scale=1.15]{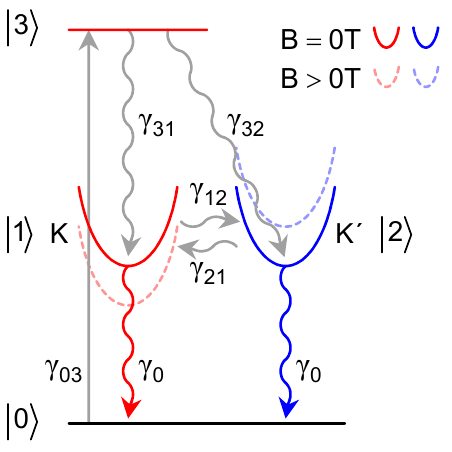}
\vspace{-10pt} \caption{Schematics of the model system: energy levels and transition rates of a four level system with the crystal ground state $\lvert 0 \rangle$, the $K$ exciton state $\lvert 1 \rangle$, the $K'$ exciton state $\lvert 2 \rangle$, and the excited $K$ exciton state $\lvert 3 \rangle$. The solid (dashed) red and blue dispersions represent $K$ and $K'$ excitons without (with) an external magnetic field. $\gamma_{03}$ denotes the absorption rate from the crystal ground state into the excited state of the $K$ valley exciton, $\gamma_{31}$ and $\gamma_{32}$ are the valley conserving and valley flipping relaxation rates, $\gamma_{12}$ and $\gamma_{21}$ are the valley flipping rates among the $K$ and $K'$ states, and $\gamma_{0}$ is the $K$ and $K'$ exciton decay rate (including both radiative and non-radiative channels).}
\label{sfig11}
\end{center}
\end{figure}

For a $\sigma^{+}$ ($\sigma^{-}$) polarized optical transition that is assumed to create population exclusively in the excited $K$ ($K'$) state, the temporal evolution of the populations in each state is given by the following set of rate equations:
\begin{equation}
\label{eq_rates}
\begin{array}{llllrcl}
\textrm{ground state}   &   \lvert 0 \rangle    & : & \quad & \dot{\rho_{0}}    &   =   & \rho_{1} \gamma_{0}  + \rho_{2} \gamma_{0}  - \rho_{0} \gamma_{03} \\
\textrm{$K$ state}      &   \lvert 1 \rangle    & : & \quad & \dot{\rho_{1}}    &   =   & \rho_{3} \gamma_{31} + \rho_{2} \gamma_{21} - \rho_{1} \gamma_{12} - \rho_{1}\gamma_{0} \\
\textrm{$K'$ state}     &   \lvert 2 \rangle    & : & \quad & \dot{\rho_{2}}    &   =   & \rho_{3} \gamma_{32} - \rho_{2} \gamma_{21} + \rho_{1} \gamma_{12} - \rho_{2}\gamma_{0} \\
\textrm{excited state}  &   \lvert 3 \rangle    & : & \quad & \dot{\rho_{3}}    &   =   & \rho_{0} \gamma_{03} - \rho_{3} \gamma_{31} - \rho_{3} \gamma_{32},
\end{array}
\end{equation}
where $\rho_{i}$ is the population of the $i$th state, $\dot{\rho_{i}}$ is the temporal derivative of $\rho_{i}$, and $\gamma_{ij}$ is the transition rate from state $i$ to $j$. The population dynamics are governed by the following rates: $\gamma_{03}$ denotes the absorption rate from the crystal ground state into an excited state in one of the two valleys, $\gamma_{31}$ and $\gamma_{32}$ are the valley conserving and valley flipping relaxation rates (or vice versa), $\gamma_{12}$ and $\gamma_{21}$ are the valley flipping rates among the $K$ and $K'$ states, and $\gamma_{0}$ is the $K$ and $K'$ exciton decay rate (including both radiative and non-radiative channels). The total population of the system is normalized to $\rho_{0} + \rho_{1} + \rho_{2} + \rho_{3} = 1$, and steady-state solutions are obtained for $\dot{\rho_{i}} = 0$.

The degree of circular polarization $P_c = (I_{co} - I_{cr})/(I_{co} + I_{cr})$ is obtained from steady-state populations $\rho_{1}$ and $\rho_{2}$ of the $K$ and $K'$ valleys:
\begin{equation}
\label{eq_pcshort}
P_{c} = \kappa \, \frac{\rho_{1}-\rho_{2}}{\rho_{1}+\rho_{2}},
\end{equation}
with the explicit expression equivalent to the one derived in the Supplementary Information of Ref.~\citenum{SI_Aivazian2015}:
\begin{equation}
\label{eq_pclong}
P_{c} = \kappa \, \left( \frac{\gamma_{0}}{\gamma_{0}+\gamma_{12}+\gamma_{21}} \cdot \frac{\gamma_{31}-\gamma_{32}}{\gamma_{31}+\gamma_{32}} \ + \
\frac{\gamma_{21}-\gamma_{12}}{\gamma_{0}+\gamma_{12}+\gamma_{21}} \right).
\end{equation}
We rewrite this expression as:
\begin{equation}
\label{eq_pclong_b}
P_{c} = \kappa \, \left(
\frac{\gamma_{0}}{\gamma_{0}+\gamma_{12}+\gamma_{21}} \cdot
\frac{1-b_{\kappa}}{1+b_{\kappa}} \ + \
\frac{\gamma_{21}-\gamma_{12}}{\gamma_{0}+\gamma_{12}+\gamma_{21}}
\right),
\end{equation}
with a branching parameter $b_{\kappa}$ given by the ratio of the valley flipping to the valley conserving relaxation rates:
\begin{equation}
\label{eq_b}
b_{\kappa} = \left( \frac{\gamma_{32}}{\gamma_{31}} \right)^{\kappa}.
\end{equation}
The definition of the branching parameter implies $b_{\kappa} =
\gamma_{32} / \gamma_{31}$ $\left( b_{\kappa} = \gamma_{31} /
\gamma_{32} \right)$ for $\kappa=+1$ ($\kappa=-1$).

\vspace{15pt}

First, we examine the the degree of circular polarization given by Eqs.~\ref{eq_pclong} and \ref{eq_pclong_b} at zero magnetic field. In the presence of time-reversal symmetry, the valley flipping processes are symmetric and thus $\gamma_{12} = \gamma_{21}$. Moreover, in the limit of ideal initial polarization, i. e. for $b_{\kappa}=0$, Eqs.~\ref{eq_pclong} and \ref{eq_pclong_b} simplify to:
\begin{equation}
\label{eq_pc0mak}
P^{0}_{c} = \frac{1}{1+2 \gamma_{l}/\gamma_{0}}=\frac{1}{1+2r_0},
\end{equation}
where $r_0=\gamma_{12}/\gamma_{0}=\gamma_{21}/\gamma_{0}$ is the ratio of the zero-field longitudinal valley depolarization rate $\gamma_{12}=\gamma_{21}=\gamma_{l}$ to the exciton decay rate $\gamma_{0}$. This expression is equivalent to the one derived in Ref.~\citenum{SI_Mak2012}, and it corresponds to the steady-state degree of circular polarization in optical spin orientation with ideal initial polarization \cite{SI_Meier1984}:
\begin{equation}
\label{eq_pc0meier}
P^{0}_{c}=\frac{1}{1 + \tau_{0}/\tau_{l}},
\end{equation}
if we identify $\tau_0 = 1/\gamma_{0}$ as the exciton lifetime and $\tau_{l} = 1/(2 \gamma_{l})$ as the valley depolarization time. This limit where $r_0$ and thus $\tau_{0}/\tau_{l}$ are fixed by the degree of circular polarization $P^{0}_{c}$ was used in the main text to discuss the data at zero magnetic field.

If the initial polarization of the fundamental exciton populations in $K$ and $K'$ valleys is non-ideal because of valley flipping events upon relaxation from the optically excited state, we have analogous to imperfect spin orientation \cite{SI_Meier1984}:
\begin{equation}
\label{eq_pc0meierpi}
P^{0}_{c}=P_i^0 \cdot \frac{1}{1+2r_0}=
\frac{1-b_{\kappa}}{1+b_{\kappa}} \cdot \frac{1}{1+2r_0},
\end{equation}
with the initial polarization $P_i^0=(1-b_{\kappa})/(1+b_{\kappa})$ given by the yield of optical valley polarization of $K$ and $K'$ populations at zero magnetic field in the presence of finite branching $b_{\kappa}>0$.

In a finite magnetic field $B$ applied perpendicular to the sample, the valley Zeeman splitting $\Delta_{\mathrm{v}}$ \cite{SI_Li2014, SI_Srivastava2015, SI_Aivazian2015, SI_MacNeill2015, SI_Wang2015a, SI_Stier2016} introduces an imbalance between $\gamma_{12}$ and $\gamma_{21}$. In the presence of non-zero branching, we return to Eqs.~\ref{eq_pclong} and \ref{eq_pclong_b} to recapitulate the findings of Ref.~\citenum{SI_Aivazian2015} with respect to the magnetic field evolution of $P_c$. First we note qualitatively that due to the valley Zeeman splitting, the imbalance of $\gamma_{12}$ and $\gamma_{21}$ increases (decreases) the population transfer rate from the upper (lower) to the lower (upper) Zeeman branch. Thus, provided sufficiently rapid valley depolarization on the timescale of the exciton lifetime, we expect the population to relax into the energetically lower valley. Consequently, the $P_c$ of the upper (lower) Zeeman valley should decrease (increase) with magnetic field. The opposite trend is observed in our study of $A$ and $L$ excitons in monolayer MoS$_2$ away from defects in agreement with the $A$ exciton response in monolayer WSe$_2$ reported in Ref.~\citenum{SI_Aivazian2015}. This counter-intuitive population distribution was attributed to (i) slow inter-valley scattering and (ii) polarization protecting branching ratio $b_{\kappa}$ that decreases (increases) for the upper (lower) valley with magnetic field \cite{SI_Aivazian2015}. Since no explicit expression for the functional form of the evolution of $b_{\kappa}$ with magnetic field was given in Ref.~\citenum{SI_Aivazian2015}, and to account for different materials and experimental conditions, we approximate $b_{\kappa}$ with a linear function in $B$:
\begin{equation}
\label{eq_bB}
b_{\kappa}(B)  = b_{0}+ \kappa \delta B,
\end{equation}
where $b_0$ is the zero-field branching ratio and $\delta$ is a proportionality factor. It is worth noting that Eq.~\ref{eq_bB} should be interpreted as a low-field approximation since in it would yield unphysical negative values for $b_{\kappa}(B)$ in the limit of sufficiently high fields. Moreover, a difference in $\delta$ for the $K$ and $K'$ valleys could account for broken time-reversal symmetry in absence of a magnetic field as pointed out in Ref.~\citenum{SI_Aivazian2015}.

To proceed with the analysis of our data using the model of Ref.~\citenum{SI_Aivazian2015} where $\gamma_{l} \ll \gamma_{0}$ (or correspondingly $r_0 \simeq 0$) was assumed, we fit the $P_c$ evolutions of the $A$ and $L$ excitons on a regular position of the triangular flake with $r_0$, $b_0$ and $\delta$ as fitting parameters. The results of best fits to the data, obtained from least $\chi^{2}_{\mathrm{red}}$ deviation (defined by the unweighted sum of squared deviations between fits and data and divided by the number of degrees of freedom), are presented in \figref{sfig12}.

\begin{figure}[!t]
\begin{center}
\includegraphics[scale=1.15]{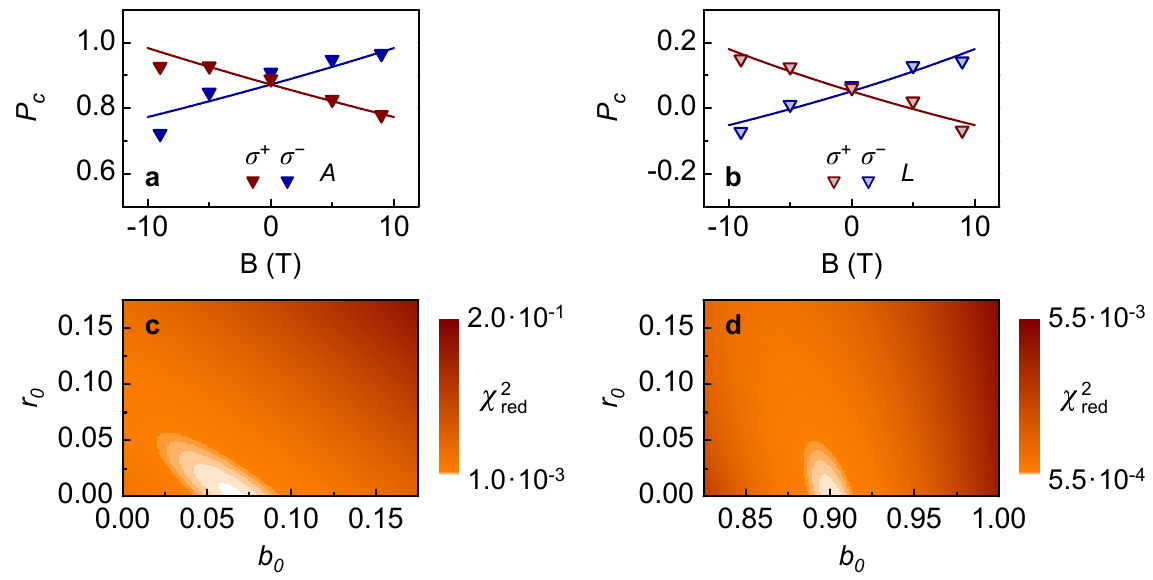}
\vspace{-10pt} \caption{\textbf{a}, \textbf{b}, $P_c$ as a function of magnetic field for $A$ and $L$ excitons; the data are reproduced from Fig.~3 of the main text. Solid lines are least $\chi^{2}_{\mathrm{red}}$ fits according to Eqs.~\ref{eq_pclong_b} and \ref{eq_bB}. Best fits were obtained with $b_0=0.07$ and $0.90$, $\delta = 0.006$~T$^{-1}$ and $0.018$~T$^{-1}$ for $A$ and $L$ excitons, respectively, and $r_0 = 0$ for both $A$ and $L$. \textbf{c}, \textbf{d}, Associated $\chi^{2}_{\mathrm{red}}$ as a function of $r_0$ and $b_0$ for $A$ and $L$, and $\delta$ values given above.}
\label{sfig12}
\end{center}
\end{figure}

The non-thermal population distribution of both $A$ and $L$ excitons is reflected by the counter-intuitive $X$-pattern of higher $P_c$ values for the Zeeman branches with higher energy. It is consistent with slow longitudinal valley depolarization on the timescale of the exciton lifetime, and best fits actually yield $r_0=0$ for both $A$ and $L$ excitons. The zero-field branching ratio for the selectively initialized population is small for $A$ excitons ($b_0=0.07$) and large for $L$ excitons ($b_0=0.90$) as expected from the less resonant excitation of the latter. In this framework, the evolution of the branching ratio with magnetic field critically determines the respective $P_c$ pattern: both $A$ and $L$ excitons exhibit an increasing protection of the optical valley polarization in the upper Zeeman valley as the branching ratio between the valley flipping and conserving relaxation rates decreases ($\delta>0$ for both $A$ and $L$) due to exchange-modified exciton dispersions \cite{SI_Aivazian2015}.

The model of Ref.~\citenum{SI_Aivazian2015} within the framework of Eqs.~\ref{eq_pclong_b} and \ref{eq_bB} requires vanishingly small values of $r_0$ without quantifying its smallness. To obtain an estimate for the ratio of the longitudinal valley depolarization time to the exciton lifetime, we extend the model of Ref.~\citenum{SI_Aivazian2015} by assigning thermal imbalances to the longitudinal inter-valley flipping rates $\gamma_{12}$ and $\gamma_{21}$ explicitly using the Boltzmann factor $\exp(\beta)$:
\begin{equation}
\label{eq_zeeman}
\begin{array}{rcl}
\gamma_{12} &   =   &   r_0 \gamma_{0}\exp(-\beta) \\
\gamma_{21} &   =   &   r_0 \gamma_{0}\exp(+\beta),
\end{array}
\end{equation}
where, at a temperature $T$, the exponent $\beta= (\Delta_{\mathrm{v}}/2)/(k_{\mathrm{B}}T)$ is determined by the Boltzmann constant $k_{\mathrm{B}}$ and the valley Zeeman splitting $\Delta_{\mathrm{v}} = g \mu_{\mathrm{B}}B$ linear in the magnetic field $B$ (with $g$ being the effective exciton $g$-factor, and $\mu_{\mathrm{B}}$ the Bohr magneton) \cite{SI_Li2014, SI_Srivastava2015, SI_Aivazian2015, SI_MacNeill2015, SI_Wang2015a, SI_Stier2016}. The zero-field ratio $r_0$, defined by the Eq.~\ref{eq_pc0mak} above, is likely to exhibit a magnetic field dependence but short of knowledge of its functional form is assumed here as constant. The exponential imbalance of $\gamma_{12}$ and $\gamma_{21}$ is responsible for an effective unidirectional thermalization of population from the upper to the lower Zeeman branch during the exciton lifetime, and modifies Eq.~\ref{eq_pclong_b} to:
\begin{equation}
\label{eq_pcbr} P_{c} = \frac{1}{1 + 2 r_0 \cosh(\beta)} \cdot
\frac{1-b_{\kappa}}{1+b_{\kappa}} \ + \ \kappa \cdot \frac{2 r_0
\sinh(\beta)}{1 + 2 r_0 \cosh(\beta)}.
\end{equation}
The degree of circular polarization is now sensitive to the thermal imbalance of the inter-valley scattering rates through the Boltzmann factor, and it accounts for the non-zero branching in the polarization initialization with its characteristic magnetic field dependence via $b_{\kappa}(B)$ given by Eq.~\ref{eq_bB}. The second term drives the thermal population distribution among the $K$ and $K'$ valleys, while the first term counteracts this thermalization via the functional dependence of the branching ratio $b_{\kappa}$ on the magnetic field in favor of a 'hot' population distribution.

\begin{figure}[!t]
\begin{center}
\includegraphics[scale=1.15]{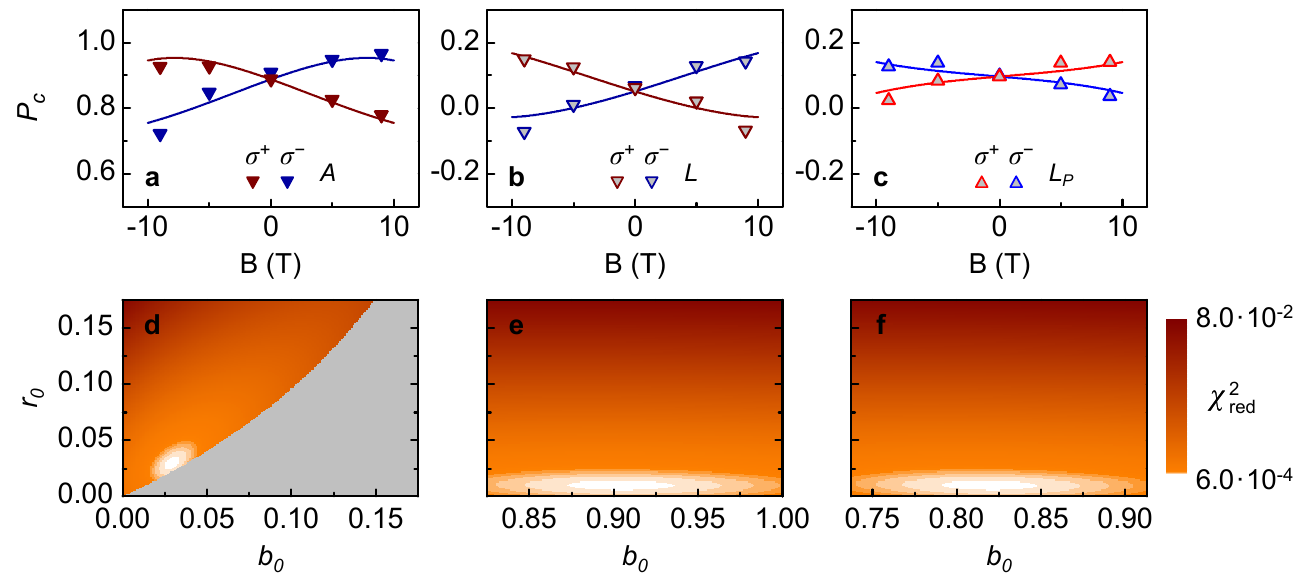}
\vspace{-10pt} \caption{\textbf{a}~-~\textbf{c}, Evolution of $A$, $L$, and $L_P$ exciton $P_c$ with magnetic field (the data is reproduced from Fig.~3 of the main text). Solid lines are a least $\chi^{2}_{\mathrm{red}}$ simultaneous fit to the data according to the model given by Eqs.~\ref{eq_bB} and \ref{eq_pcbr}. To reduce the number of free fitting parameters, the $g$-factor of the $A$ excitons was fixed to $g=4$, and a global temperature $T$ was assumed. Moreover, $\delta=0$ was used for $L_P$, and $g$ and $r_0$ were set equal for $L$ and $L_P$. Best fit was obtained for the following values of the remaining free fitting parameters: $b_0 = 0.03$, $0.90$, and $0.82$ for $A$, $L$, and $L_P$; $r_0 = 0.03$ for $A$ and $0.01$ for $L$ and $L_P$; $\delta = 0.012$~T$^{-1}$ for $A$ and $0.027$~T$^{-1}$ for $L$; $g=4.4$ for $L$ and $L_P$; $T=9$~K. \textbf{d}~-~\textbf{f}, Corresponding $\chi^{2}_{\mathrm{red}}$ values as a function of $r_0$ and $b_0$ for $A$, $L$, and $L_P$ with other parameters as given above. The gray region in \textbf{d} with unphysical values of $\lvert P_c \rvert > 1$ was not evaluated.}
\label{sfig13}
\end{center}
\end{figure}

With this expression for the field-evolution of $P_c$, we performed a simultaneous fit of our set of data for $A$ and $L$ excitons away from the puddle and $L_P$ excitons. To minimize the number of free fitting parameters, we assumed no field-dependence of $b_{\kappa}$ for the case of $L_P$ excitons that exhibit a thermal population distribution, and set both the $g$-factors and the ratios $r_0$ equal for both $L$ and $L_P$. Moreover, we fixed the $A$ exciton $g$-factor to $4.0$ \cite{SI_Wang2015a,SI_Stier2016}, and assumed a global temperature $T$ for the entire set of data. With these parameters fixed, best simultaneous fit was obtained with $b_0 = 0.03$, $0.90$, and $0.82$ for $A$, $L$, and $L_P$; $r_0 = 0.03$ for $A$ and $0.01$ for $L$ and $L_P$; $\delta = 0.012$~T$^{-1}$ for $A$ and $0.027$~T$^{-1}$ for $L$; $g=4.4$ for $L$ and $L_P$; $T=9$~K. The corresponding evolutions of $P_c$ for $A$, $L$ and $L_P$ excitons are shown in \figref{sfig13}a,~b and c, respectively. We note that the best fit yielded an effective temperature of $T=9$~K consistent with local laser heating, and $g=4.4$ for $L$ and $L_P$ excitons in agreement with elevated values reported for excitons localized in quantum dots \cite{SI_Srivastava2015a,SI_He2015,SI_Koperski2015,SI_Chakraborty2015}.

The main effect of the model of Eq.~\ref{eq_pcbr} as compared to the model of Eq.~\ref{eq_pclong_b} is the increased field dependence of the $A$ exciton branching ($\delta$ increased by a factor of two from $0.006$~T$^{-1}$ to $0.012$~T$^{-1}$), whereas the values of the zero-field branching ratio of $b_0$ for the valley flipping to the valley conserving rate are comparable for the two models. Most importantly, the model yields an estimate for $r_0$ for both $A$ and localized excitons: bet fit values of $r_0=0.03$ for $A$, and $0.01$ for $L$ and $L_P$ correspond to $\tau_l/ \tau_0 \simeq 17$ and $50$, respectively. Taking the PL decay time of $4.5$~ps for the $A$ excitons in MoS$_2$ \cite{SI_Lagarde2014} that we do not resolve in our time-correlated PL experiments, we thus obtain an estimate for the longitudinal valley depolarization time of $\tau_l \simeq 80$~ps for $A$ excitons. With the same scaling we estimate $\tau_l \simeq 230$~ps for localized excitons that exhibit PL decay dynamics dominated by the fast component below the resolution limit of our fiber-based setup with dispersion (as shown in \figref{sfig3}b, only $\sim 10\%$ of the total PL intensity of both $L$ and $L_P$ excitons contribute to the slow decay component). The main difference for $L$ and $L_p$ populations stems from different branching during relaxation: while the upper valley polarization of $L$ excitons is weakly yet increasingly protected with magnetic field (with a slope of $0.027$~T$^{-1}$), the branching ratio of $L_P$ excitons is independent of magnetic field. Given similar PL decay dynamics in the $L$ and $L_P$ bands, and similar branching ratios at zero magnetic field, we conclude that hot and thermal valley populations of localized excitons arise from field-dependent and field-independent branching ratios, respectively.



\end{supplement}


\begin{thebibliography}{10}
\expandafter\ifx\csname url\endcsname\relax
  \def\url#1{\texttt{#1}}\fi
\expandafter\ifx\csname urlprefix\endcsname\relax\def\urlprefix{URL }\fi
\providecommand{\bibinfo}[2]{#2}
\providecommand{\eprint}[2][]{\url{#2}}

\bibitem{Gunawan2007}
\bibinfo{author}{Gunawan, O.}, \bibinfo{author}{De~Poortere, E.~P.} \&
  \bibinfo{author}{Shayegan, M.}
\newblock \bibinfo{title}{{AlAs} two-dimensional electrons in an antidot
  lattice: Electron pinball with elliptical {Fermi} contours}.
\newblock \emph{\bibinfo{journal}{Phys. Rev. B}} \textbf{\bibinfo{volume}{75}},
  \bibinfo{pages}{081304} (\bibinfo{year}{2007}).

\bibitem{Culcer2012}
\bibinfo{author}{Culcer, D.}, \bibinfo{author}{Saraiva, A.~L.},
  \bibinfo{author}{Koiller, B.}, \bibinfo{author}{Hu, X.} \&
  \bibinfo{author}{Das~Sarma, S.}
\newblock \bibinfo{title}{Valley-based noise-resistant quantum computation
  using {Si} quantum dots}.
\newblock \emph{\bibinfo{journal}{Phys. Rev. Lett.}}
  \textbf{\bibinfo{volume}{108}}, \bibinfo{pages}{126804}
  (\bibinfo{year}{2012}).

\bibitem{Rycerz2007}
\bibinfo{author}{Rycerz, A.}, \bibinfo{author}{Tworzyd{\l}o, J.} \&
  \bibinfo{author}{Beenakker, C. W.~J.}
\newblock \bibinfo{title}{Valley filter and valley valve in graphene}.
\newblock \emph{\bibinfo{journal}{Nat. Phys.}} \textbf{\bibinfo{volume}{3}},
  \bibinfo{pages}{172--175} (\bibinfo{year}{2007}).

\bibitem{Xiao2012}
\bibinfo{author}{Xiao, D.}, \bibinfo{author}{Liu, G.-B.},
  \bibinfo{author}{Feng, W.}, \bibinfo{author}{Xu, X.} \& \bibinfo{author}{Yao,
  W.}
\newblock \bibinfo{title}{Coupled spin and valley physics in monolayers of
  {MoS$_{2}$} and other group-{VI} dichalcogenides}.
\newblock \emph{\bibinfo{journal}{Phys. Rev. Lett.}}
  \textbf{\bibinfo{volume}{108}}, \bibinfo{pages}{196802}
  (\bibinfo{year}{2012}).

\bibitem{Xu2014}
\bibinfo{author}{Xu, X.}, \bibinfo{author}{Yao, W.}, \bibinfo{author}{Xiao, D.}
  \& \bibinfo{author}{Heinz, T.~F.}
\newblock \bibinfo{title}{Spin and pseudospins in layered transition metal
  dichalcogenides}.
\newblock \emph{\bibinfo{journal}{Nat. Phys.}} \textbf{\bibinfo{volume}{10}},
  \bibinfo{pages}{343--350} (\bibinfo{year}{2014}).

\bibitem{Splendiani2010}
\bibinfo{author}{Splendiani, A.} \emph{et~al.}
\newblock \bibinfo{title}{Emerging photoluminescence in monolayer {MoS$_{2}$}}.
\newblock \emph{\bibinfo{journal}{Nano Lett.}} \textbf{\bibinfo{volume}{10}},
  \bibinfo{pages}{1271--1275} (\bibinfo{year}{2010}).

\bibitem{Mak2010}
\bibinfo{author}{Mak, K.~F.}, \bibinfo{author}{Lee, C.}, \bibinfo{author}{Hone,
  J.}, \bibinfo{author}{Shan, J.} \& \bibinfo{author}{Heinz, T.~F.}
\newblock \bibinfo{title}{Atomically thin {MoS$_{2}$}: A new direct-gap
  semiconductor}.
\newblock \emph{\bibinfo{journal}{Phys. Rev. Lett.}}
  \textbf{\bibinfo{volume}{105}}, \bibinfo{pages}{136805}
  (\bibinfo{year}{2010}).

\bibitem{Kim2014}
\bibinfo{author}{Kim, J.} \emph{et~al.}
\newblock \bibinfo{title}{Ultrafast generation of pseudo-magnetic field for
  valley excitons in {WSe$_{2}$} monolayers}.
\newblock \emph{\bibinfo{journal}{Science}} \textbf{\bibinfo{volume}{346}},
  \bibinfo{pages}{1205--1208} (\bibinfo{year}{2014}).

\bibitem{Cao2012}
\bibinfo{author}{Cao, T.} \emph{et~al.}
\newblock \bibinfo{title}{Valley-selective circular dichroism of monolayer
  molybdenum disulphide}.
\newblock \emph{\bibinfo{journal}{Nat. Commun.}} \textbf{\bibinfo{volume}{3}},
  \bibinfo{pages}{887} (\bibinfo{year}{2012}).

\bibitem{Mak2012}
\bibinfo{author}{Mak, K.~F.}, \bibinfo{author}{He, K.}, \bibinfo{author}{Shan,
  J.} \& \bibinfo{author}{Heinz, T.~F.}
\newblock \bibinfo{title}{Control of valley polarization in monolayer
  {MoS$_{2}$} by optical helicity}.
\newblock \emph{\bibinfo{journal}{Nat. Nanotechnol.}}
  \textbf{\bibinfo{volume}{7}}, \bibinfo{pages}{494--498}
  (\bibinfo{year}{2012}).

\bibitem{Zeng2012}
\bibinfo{author}{Zeng, H.}, \bibinfo{author}{Dai, J.}, \bibinfo{author}{Yao,
  W.}, \bibinfo{author}{Xiao, D.} \& \bibinfo{author}{Cui, X.}
\newblock \bibinfo{title}{Valley polarization in {MoS$_{2}$} monolayers by
  optical pumping}.
\newblock \emph{\bibinfo{journal}{Nat. Nanotechnol.}}
  \textbf{\bibinfo{volume}{7}}, \bibinfo{pages}{490--493}
  (\bibinfo{year}{2012}).

\bibitem{Zhang2014}
\bibinfo{author}{Zhang, Y.~J.}, \bibinfo{author}{Oka, T.},
  \bibinfo{author}{Suzuki, R.}, \bibinfo{author}{Ye, J.~T.} \&
  \bibinfo{author}{Iwasa, Y.}
\newblock \bibinfo{title}{Electrically switchable chiral light-emitting
  transistor}.
\newblock \emph{\bibinfo{journal}{Science}} \textbf{\bibinfo{volume}{344}},
  \bibinfo{pages}{725--728} (\bibinfo{year}{2014}).

\bibitem{Mak2014}
\bibinfo{author}{Mak, K.~F.}, \bibinfo{author}{McGill, K.~L.},
  \bibinfo{author}{Park, J.} \& \bibinfo{author}{McEuen, P.~L.}
\newblock \bibinfo{title}{The valley {Hall} effect in {MoS$_{2}$} transistors}.
\newblock \emph{\bibinfo{journal}{Science}} \textbf{\bibinfo{volume}{344}},
  \bibinfo{pages}{1489--1492} (\bibinfo{year}{2014}).

\bibitem{Sallen2012}
\bibinfo{author}{Sallen, G.} \emph{et~al.}
\newblock \bibinfo{title}{Robust optical emission polarization in {MoS$_{2}$}
  monolayers through selective valley excitation}.
\newblock \emph{\bibinfo{journal}{Phys. Rev. B}} \textbf{\bibinfo{volume}{86}},
  \bibinfo{pages}{081301} (\bibinfo{year}{2012}).

\bibitem{Lagarde2014}
\bibinfo{author}{Lagarde, D.} \emph{et~al.}
\newblock \bibinfo{title}{Carrier and polarization dynamics in monolayer
  {MoS$_{2}$}}.
\newblock \emph{\bibinfo{journal}{Phys. Rev. Lett.}}
  \textbf{\bibinfo{volume}{112}}, \bibinfo{pages}{047401}
  (\bibinfo{year}{2014}).

\bibitem{Jones2013}
\bibinfo{author}{Jones, A.~M.} \emph{et~al.}
\newblock \bibinfo{title}{Optical generation of excitonic valley coherence in
  monolayer {WSe$_{2}$}}.
\newblock \emph{\bibinfo{journal}{Nat. Nanotechnol.}}
  \textbf{\bibinfo{volume}{8}}, \bibinfo{pages}{634--638}
  (\bibinfo{year}{2013}).

\bibitem{Wang2014}
\bibinfo{author}{Wang, G.} \emph{et~al.}
\newblock \bibinfo{title}{Valley dynamics probed through charged and neutral
  exciton emission in monolayer {WSe$_{2}$}}.
\newblock \emph{\bibinfo{journal}{Phys. Rev. B}} \textbf{\bibinfo{volume}{90}},
  \bibinfo{pages}{075413} (\bibinfo{year}{2014}).

\bibitem{Wu2013}
\bibinfo{author}{Wu, S.} \emph{et~al.}
\newblock \bibinfo{title}{Electrical tuning of valley magnetic moment through
  symmetry control in bilayer {MoS$_{2}$}}.
\newblock \emph{\bibinfo{journal}{Nat. Phys.}} \textbf{\bibinfo{volume}{9}},
  \bibinfo{pages}{149--153} (\bibinfo{year}{2013}).

\bibitem{Jones2014}
\bibinfo{author}{Jones, A.~M.} \emph{et~al.}
\newblock \bibinfo{title}{Spin-layer locking effects in optical orientation of
  exciton spin in bilayer {WSe$_{2}$}}.
\newblock \emph{\bibinfo{journal}{Nat. Phys.}} \textbf{\bibinfo{volume}{10}},
  \bibinfo{pages}{130--134} (\bibinfo{year}{2014}).

\bibitem{Zhu2014}
\bibinfo{author}{Zhu, B.}, \bibinfo{author}{Zeng, H.}, \bibinfo{author}{Dai,
  J.}, \bibinfo{author}{Gong, Z.} \& \bibinfo{author}{Cui, X.}
\newblock \bibinfo{title}{Anomalously robust valley polarization and valley
  coherence in bilayer {WS$_{2}$}}.
\newblock \emph{\bibinfo{journal}{Proc. Natl. Acad. Sci. USA}}
  \textbf{\bibinfo{volume}{111}}, \bibinfo{pages}{11606--11611}
  (\bibinfo{year}{2014}).

\bibitem{Meier1984}
\bibinfo{editor}{Meier, F.} \& \bibinfo{editor}{Zakharchenya, B.~P.} (eds.)
  \emph{\bibinfo{title}{Optical Orientation}} (\bibinfo{publisher}{Elsevier
  Science Publishers B.V.}, \bibinfo{year}{1984}).

\bibitem{Maialle1993}
\bibinfo{author}{Maialle, M.~Z.}, \bibinfo{author}{de~Andrada~e Silva, E.~A.}
  \& \bibinfo{author}{Sham, L.~J.}
\newblock \bibinfo{title}{Exciton spin dynamics in quantum wells}.
\newblock \emph{\bibinfo{journal}{Phys. Rev. B}} \textbf{\bibinfo{volume}{47}},
  \bibinfo{pages}{15776--15788} (\bibinfo{year}{1993}).

\bibitem{Glazov2014}
\bibinfo{author}{Glazov, M.~M.} \emph{et~al.}
\newblock \bibinfo{title}{Exciton fine structure and spin decoherence in
  monolayers of transition metal dichalcogenides}.
\newblock \emph{\bibinfo{journal}{Phys. Rev. B}} \textbf{\bibinfo{volume}{89}},
  \bibinfo{pages}{201302} (\bibinfo{year}{2014}).

\bibitem{Yu2014a}
\bibinfo{author}{Yu, T.} \& \bibinfo{author}{Wu, M.~W.}
\newblock \bibinfo{title}{Valley depolarization due to intervalley and
  intravalley electron-hole exchange interactions in monolayer {MoS$_{2}$}}.
\newblock \emph{\bibinfo{journal}{Phys. Rev. B}} \textbf{\bibinfo{volume}{89}},
  \bibinfo{pages}{205303} (\bibinfo{year}{2014}).

\bibitem{Tongay2013}
\bibinfo{author}{Tongay, S.} \emph{et~al.}
\newblock \bibinfo{title}{Defects activated photoluminescence in
  two-dimensional semiconductors: interplay between bound, charged, and free
  excitons}.
\newblock \emph{\bibinfo{journal}{Sci. Rep.}} \textbf{\bibinfo{volume}{3}},
  \bibinfo{pages}{2657} (\bibinfo{year}{2013}).

\bibitem{Zande2013}
\bibinfo{author}{Van~der Zande, A.~M.} \emph{et~al.}
\newblock \bibinfo{title}{Grains and grain boundaries in highly crystalline
  monolayer molybdenum disulphide}.
\newblock \emph{\bibinfo{journal}{Nat. Mater.}} \textbf{\bibinfo{volume}{12}},
  \bibinfo{pages}{554--561} (\bibinfo{year}{2013}).

\bibitem{Najmaei2013}
\bibinfo{author}{Najmaei, S.} \emph{et~al.}
\newblock \bibinfo{title}{Vapour phase growth and grain boundary structure of
  molybdenum disulphide atomic layers}.
\newblock \emph{\bibinfo{journal}{Nat. Mater.}} \textbf{\bibinfo{volume}{12}},
  \bibinfo{pages}{754--759} (\bibinfo{year}{2013}).

\bibitem{Srivastava2015a}
\bibinfo{author}{Srivastava, A.} \emph{et~al.}
\newblock \bibinfo{title}{Optically active quantum dots in monolayer
  {WSe$_{2}$}}.
\newblock \emph{\bibinfo{journal}{Nat. Nanotechnol.}}
  \textbf{\bibinfo{volume}{10}}, \bibinfo{pages}{491--496}
  (\bibinfo{year}{2015}).

\bibitem{He2015}
\bibinfo{author}{He, Y.-M.} \emph{et~al.}
\newblock \bibinfo{title}{Single quantum emitters in monolayer semiconductors}.
\newblock \emph{\bibinfo{journal}{Nat. Nanotechnol.}}
  \textbf{\bibinfo{volume}{10}}, \bibinfo{pages}{497--502}
  (\bibinfo{year}{2015}).

\bibitem{Koperski2015}
\bibinfo{author}{Koperski, M.} \emph{et~al.}
\newblock \bibinfo{title}{Single photon emitters in exfoliated {WSe$_{2}$}
  structures}.
\newblock \emph{\bibinfo{journal}{Nat. Nanotechnol.}}
  \textbf{\bibinfo{volume}{10}}, \bibinfo{pages}{503--506}
  (\bibinfo{year}{2015}).

\bibitem{Chakraborty2015}
\bibinfo{author}{Chakraborty, C.}, \bibinfo{author}{Kinnischtzke, L.},
  \bibinfo{author}{Goodfellow, K.~M.}, \bibinfo{author}{Beams, R.} \&
  \bibinfo{author}{Vamivakas, A.~N.}
\newblock \bibinfo{title}{Voltage-controlled quantum light from an atomically
  thin semiconductor}.
\newblock \emph{\bibinfo{journal}{Nat. Nanotechnol.}}
  \textbf{\bibinfo{volume}{10}}, \bibinfo{pages}{507--511}
  (\bibinfo{year}{2015}).

\bibitem{Wang2015}
\bibinfo{author}{Wang, G.} \emph{et~al.}
\newblock \bibinfo{title}{Giant enhancement of the optical second-harmonic
  emission of {WSe$_{2}$} monolayers by laser excitation at exciton
  resonances}.
\newblock \emph{\bibinfo{journal}{Phys. Rev. Lett.}}
  \textbf{\bibinfo{volume}{114}}, \bibinfo{pages}{097403}
  (\bibinfo{year}{2015}).

\bibitem{Gong2013}
\bibinfo{author}{Gong, Z.} \emph{et~al.}
\newblock \bibinfo{title}{Magnetoelectric effects and valley-controlled spin
  quantum gates in transition metal dichalcogenide bilayers}.
\newblock \emph{\bibinfo{journal}{Nat. Commun.}} \textbf{\bibinfo{volume}{4}},
  \bibinfo{pages}{2053} (\bibinfo{year}{2013}).

\bibitem{Li2014}
\bibinfo{author}{Li, Y.} \emph{et~al.}
\newblock \bibinfo{title}{Valley splitting and polarization by the {Zeeman}
  effect in monolayer {MoSe$_{2}$}}.
\newblock \emph{\bibinfo{journal}{Phys. Rev. Lett.}}
  \textbf{\bibinfo{volume}{113}}, \bibinfo{pages}{266804}
  (\bibinfo{year}{2014}).

\bibitem{Srivastava2015}
\bibinfo{author}{Srivastava, A.} \emph{et~al.}
\newblock \bibinfo{title}{Valley {Zeeman} effect in elementary optical
  excitations of monolayer {WSe$_{2}$}}.
\newblock \emph{\bibinfo{journal}{Nat. Phys.}} \textbf{\bibinfo{volume}{11}},
  \bibinfo{pages}{141--147} (\bibinfo{year}{2015}).

\bibitem{Aivazian2015}
\bibinfo{author}{Aivazian, G.} \emph{et~al.}
\newblock \bibinfo{title}{Magnetic control of valley pseudospin in monolayer
  {WSe$_{2}$}}.
\newblock \emph{\bibinfo{journal}{Nat. Phys.}} \textbf{\bibinfo{volume}{11}},
  \bibinfo{pages}{148--152} (\bibinfo{year}{2015}).

\bibitem{MacNeill2015}
\bibinfo{author}{MacNeill, D.} \emph{et~al.}
\newblock \bibinfo{title}{Breaking of valley degeneracy by magnetic field in
  monolayer {MoSe$_{2}$}}.
\newblock \emph{\bibinfo{journal}{Phys. Rev. Lett.}}
  \textbf{\bibinfo{volume}{114}}, \bibinfo{pages}{037401}
  (\bibinfo{year}{2015}).

\bibitem{Wang2015a}
\bibinfo{author}{Wang, G.} \emph{et~al.}
\newblock \bibinfo{title}{Magneto-optics in transition metal diselenide
  monolayers}.
\newblock \emph{\bibinfo{journal}{2D Mater.}} \textbf{\bibinfo{volume}{2}},
  \bibinfo{pages}{034002} (\bibinfo{year}{2015}).

\bibitem{Stier2016}
\bibinfo{author}{Stier, A.~V.}, \bibinfo{author}{McCreary, K.~M.},
  \bibinfo{author}{Jonker, B.~T.}, \bibinfo{author}{Kono, J.} \&
  \bibinfo{author}{Crooker, S.~A.}
\newblock \bibinfo{title}{Exciton diamagnetic shifts and valley {Zeeman}
  effects in monolayer {WS$_{2}$} and {MoS$_{2}$} to 65 {Tesla}}.
\newblock \emph{\bibinfo{journal}{Nat. Commun.}} \textbf{\bibinfo{volume}{7}},
  \bibinfo{pages}{10643} (\bibinfo{year}{2016}).

\bibitem{Cadiz2016}
\bibinfo{author}{Cadiz, F.} \emph{et~al.}
\newblock \bibinfo{title}{Well separated trion and neutral excitons on
  superacid treated {MoS$_{2}$} monolayers}.
\newblock \emph{\bibinfo{journal}{Appl. Phys. Lett.}}
  \textbf{\bibinfo{volume}{108}}, \bibinfo{pages}{251106}
  (\bibinfo{year}{2016}).

\bibitem{Karzig2015}
\bibinfo{author}{Karzig, T.}, \bibinfo{author}{Bardyn, C.-E.},
  \bibinfo{author}{Lindner, N.~H.} \& \bibinfo{author}{Refael, G.}
\newblock \bibinfo{title}{Topological polaritons}.
\newblock \emph{\bibinfo{journal}{Phys. Rev. X}} \textbf{\bibinfo{volume}{5}},
  \bibinfo{pages}{031001} (\bibinfo{year}{2015}).

\end{thebibliography}

\begin{thebibliography}{10}
\expandafter\ifx\csname url\endcsname\relax
  \def\url#1{\texttt{#1}}\fi
\expandafter\ifx\csname urlprefix\endcsname\relax\def\urlprefix{URL }\fi
\providecommand{\bibinfo}[2]{#2}
\providecommand{\eprint}[2][]{\url{#2}}

\bibitem{SI_Najmaei2013}
\bibinfo{author}{Najmaei, S.} \emph{et~al.}
\newblock \bibinfo{title}{Vapour phase growth and grain boundary structure of
  molybdenum disulphide atomic layers}.
\newblock \emph{\bibinfo{journal}{Nat. Mater.}} \textbf{\bibinfo{volume}{12}},
  \bibinfo{pages}{754--759} (\bibinfo{year}{2013}).

\bibitem{SI_Reina2008}
\bibinfo{author}{Reina, A.} \emph{et~al.}
\newblock \bibinfo{title}{Transferring and identification of single- and
  few-layer graphene on arbitrary substrates}.
\newblock \emph{\bibinfo{journal}{J. Phys. Chem. C}}
  \textbf{\bibinfo{volume}{112}}, \bibinfo{pages}{17741--17744}
  (\bibinfo{year}{2008}).

\bibitem{SI_Scheuschner2014}
\bibinfo{author}{Scheuschner, N.} \emph{et~al.}
\newblock \bibinfo{title}{Photoluminescence of freestanding single- and
  few-layer {MoS$_{2}$}}.
\newblock \emph{\bibinfo{journal}{Phys. Rev. B}} \textbf{\bibinfo{volume}{89}},
  \bibinfo{pages}{125406} (\bibinfo{year}{2014}).

\bibitem{SI_Berkelbach2013}
\bibinfo{author}{Berkelbach, T.~C.}, \bibinfo{author}{Hybertsen, M.~S.} \&
  \bibinfo{author}{Reichman, D.~R.}
\newblock \bibinfo{title}{Theory of neutral and charged excitons in monolayer
  transition metal dichalcogenides}.
\newblock \emph{\bibinfo{journal}{Phys. Rev. B}} \textbf{\bibinfo{volume}{88}},
  \bibinfo{pages}{045318} (\bibinfo{year}{2013}).

\bibitem{SI_Golasa2014}
\bibinfo{author}{Go{\l}asa, K.} \emph{et~al.}
\newblock \bibinfo{title}{Multiphonon resonant {Raman} scattering in
  {MoS$_{2}$}}.
\newblock \emph{\bibinfo{journal}{Appl. Phys. Lett.}}
  \textbf{\bibinfo{volume}{104}}, \bibinfo{pages}{092106}
  (\bibinfo{year}{2014}).

\bibitem{SI_Staiger2015}
\bibinfo{author}{Staiger, M.} \emph{et~al.}
\newblock \bibinfo{title}{Splitting of monolayer out-of-plane {$A_{1}^{'}$}
  {Raman} mode in few-layer {WS$_{2}$}}.
\newblock \emph{\bibinfo{journal}{Phys. Rev. B}} \textbf{\bibinfo{volume}{91}},
  \bibinfo{pages}{195419} (\bibinfo{year}{2015}).

\bibitem{SI_Lagarde2014}
\bibinfo{author}{Lagarde, D.} \emph{et~al.}
\newblock \bibinfo{title}{Carrier and polarization dynamics in monolayer
  {MoS$_{2}$}}.
\newblock \emph{\bibinfo{journal}{Phys. Rev. Lett.}}
  \textbf{\bibinfo{volume}{112}}, \bibinfo{pages}{047401}
  (\bibinfo{year}{2014}).

\bibitem{SI_Aivazian2015}
\bibinfo{author}{Aivazian, G.} \emph{et~al.}
\newblock \bibinfo{title}{Magnetic control of valley pseudospin in monolayer
  {WSe$_{2}$}}.
\newblock \emph{\bibinfo{journal}{Nat. Phys.}} \textbf{\bibinfo{volume}{11}},
  \bibinfo{pages}{148--152} (\bibinfo{year}{2015}).

\bibitem{SI_Mak2012}
\bibinfo{author}{Mak, K.~F.}, \bibinfo{author}{He, K.}, \bibinfo{author}{Shan,
  J.} \& \bibinfo{author}{Heinz, T.~F.}
\newblock \bibinfo{title}{Control of valley polarization in monolayer
  {MoS$_{2}$} by optical helicity}.
\newblock \emph{\bibinfo{journal}{Nat. Nanotechnol.}}
  \textbf{\bibinfo{volume}{7}}, \bibinfo{pages}{494--498}
  (\bibinfo{year}{2012}).

\bibitem{SI_Meier1984}
\bibinfo{editor}{Meier, F.} \& \bibinfo{editor}{Zakharchenya, B.~P.} (eds.)
  \emph{\bibinfo{title}{Optical Orientation}} (\bibinfo{publisher}{Elsevier
  Science Publishers B.V.}, \bibinfo{year}{1984}).

\bibitem{SI_Li2014}
\bibinfo{author}{Li, Y.} \emph{et~al.}
\newblock \bibinfo{title}{Valley splitting and polarization by the {Zeeman}
  effect in monolayer {MoSe$_{2}$}}.
\newblock \emph{\bibinfo{journal}{Phys. Rev. Lett.}}
  \textbf{\bibinfo{volume}{113}}, \bibinfo{pages}{266804}
  (\bibinfo{year}{2014}).

\bibitem{SI_Srivastava2015}
\bibinfo{author}{Srivastava, A.} \emph{et~al.}
\newblock \bibinfo{title}{Valley {Zeeman} effect in elementary optical
  excitations of monolayer {WSe$_{2}$}}.
\newblock \emph{\bibinfo{journal}{Nat. Phys.}} \textbf{\bibinfo{volume}{11}},
  \bibinfo{pages}{141--147} (\bibinfo{year}{2015}).

\bibitem{SI_MacNeill2015}
\bibinfo{author}{MacNeill, D.} \emph{et~al.}
\newblock \bibinfo{title}{Breaking of valley degeneracy by magnetic field in
  monolayer {MoSe$_{2}$}}.
\newblock \emph{\bibinfo{journal}{Phys. Rev. Lett.}}
  \textbf{\bibinfo{volume}{114}}, \bibinfo{pages}{037401}
  (\bibinfo{year}{2015}).

\bibitem{SI_Wang2015a}
\bibinfo{author}{Wang, G.} \emph{et~al.}
\newblock \bibinfo{title}{Magneto-optics in transition metal diselenide
  monolayers}.
\newblock \emph{\bibinfo{journal}{2D Mater.}} \textbf{\bibinfo{volume}{2}},
  \bibinfo{pages}{034002} (\bibinfo{year}{2015}).

\bibitem{SI_Stier2016}
\bibinfo{author}{Stier, A.~V.}, \bibinfo{author}{McCreary, K.~M.},
  \bibinfo{author}{Jonker, B.~T.}, \bibinfo{author}{Kono, J.} \&
  \bibinfo{author}{Crooker, S.~A.}
\newblock \bibinfo{title}{Exciton diamagnetic shifts and valley {Zeeman}
  effects in monolayer {WS$_{2}$} and {MoS$_{2}$} to 65 {Tesla}}.
\newblock \emph{\bibinfo{journal}{Nat. Commun.}} \textbf{\bibinfo{volume}{7}},
  \bibinfo{pages}{10643} (\bibinfo{year}{2016}).

\bibitem{SI_Srivastava2015a}
\bibinfo{author}{Srivastava, A.} \emph{et~al.}
\newblock \bibinfo{title}{Optically active quantum dots in monolayer
  {WSe$_{2}$}}.
\newblock \emph{\bibinfo{journal}{Nat. Nanotechnol.}}
  \textbf{\bibinfo{volume}{10}}, \bibinfo{pages}{491--496}
  (\bibinfo{year}{2015}).

\bibitem{SI_He2015}
\bibinfo{author}{He, Y.-M.} \emph{et~al.}
\newblock \bibinfo{title}{Single quantum emitters in monolayer semiconductors}.
\newblock \emph{\bibinfo{journal}{Nat. Nanotechnol.}}
  \textbf{\bibinfo{volume}{10}}, \bibinfo{pages}{497--502}
  (\bibinfo{year}{2015}).

\bibitem{SI_Koperski2015}
\bibinfo{author}{Koperski, M.} \emph{et~al.}
\newblock \bibinfo{title}{Single photon emitters in exfoliated {WSe$_{2}$}
  structures}.
\newblock \emph{\bibinfo{journal}{Nat. Nanotechnol.}}
  \textbf{\bibinfo{volume}{10}}, \bibinfo{pages}{503--506}
  (\bibinfo{year}{2015}).

\bibitem{SI_Chakraborty2015}
\bibinfo{author}{Chakraborty, C.}, \bibinfo{author}{Kinnischtzke, L.},
  \bibinfo{author}{Goodfellow, K.~M.}, \bibinfo{author}{Beams, R.} \&
  \bibinfo{author}{Vamivakas, A.~N.}
\newblock \bibinfo{title}{Voltage-controlled quantum light from an atomically
  thin semiconductor}.
\newblock \emph{\bibinfo{journal}{Nat. Nanotechnol.}}
  \textbf{\bibinfo{volume}{10}}, \bibinfo{pages}{507--511}
  (\bibinfo{year}{2015}).

\end{thebibliography}
\end{document}